\documentclass[submission,copyright,creativecommons]{eptcs}
\usepackage[english]{babel}
\usepackage{graphicx}

\providecommand{\event}{TFPIE 2014} 
\usepackage{breakurl}             

\usepackage{color}
\usepackage{listings}
\usepackage{textcomp}
\usepackage[caption=false]{subfig}

\usepackage{fancybox}
\usepackage{fancyhdr}

\usepackage{booktabs}

\usepackage{lipsum}
\usepackage{verbatim}

\newcommand{\bricklayer}{\ensuremath{\mathcal{B}\emph{ricklayer}}}

\definecolor{keywordColor}{rgb}{0.5,0.00,0.25}
\definecolor{commentColor}{rgb}{0.00,0.50,0.00}
\definecolor{stringColor}{rgb}{0.00,0.00,1.00}
\definecolor{bgColor}{rgb}{0.85,0.95,1.00}

\def\titlerunning{Bricklayer: An Authentic Introduction to the Functional Programming Language SML}
\def\authorrunning{Victor Winter}

\begin{document}

\title{\bricklayer: An Authentic Introduction to the Functional Programming Language SML}

\author{Victor Winter
\institute{Department of Computer Science\\
University of Nebraska at Omaha\\
Omaha, USA}
\email{vwinter@unomaha.edu}
}

\maketitle
\begin{abstract}
Functional programming languages are seen by many as instrumental to effectively utilizing the computational power of multi-core platforms. As a result, there is growing interest to introduce functional programming and functional thinking as early as possible within the computer science curriculum. \bricklayer\ is an API, written in SML, that provides a set of abstractions for creating LEGO\textregistered\ artifacts which can be viewed using LEGO Digital Designer. The goal of \bricklayer\ is to create a problem space (i.e., a set of LEGO artifacts) that is accessible and engaging to programmers (especially novice programmers) while providing an authentic introduction to the functional programming language SML.
\end{abstract}


\section{Motivation}

Technological advances are giving rise to computational environments in which concurrent and parallel computation are rapidly gaining prominence. This has led to a renewed interest in functional programming languages. Properties such as \emph{immutability} (i.e., statelessness) and \emph{referential transparency} make functional programs well-suited for multi-core architectures because  programs having such properties avoid problems arising from shared state (e.g., deadlock and starvation). Joe Armstrong has stated that, subject to minor caveats, ``Your Erlang program should just run N times faster on an N core processor''.\cite{Armstrong:Erlang}

The suitability of functional programming and functional thinking to multi-core architectures is creating a desire to increase the footprint of functional programming within the educational system. Spearheading the effort to bring about this educational shift is Carnegie Mellon University (CMU) which, in 2011, began teaching functional programming (using SML) to its freshmen. What is perhaps more remarkable is that, at CMU, OO programming is eliminated from its introductory curriculum.

\begin{quote}
\emph{Object-oriented programming is eliminated entirely from the introductory curriculum, because it is both anti-modular and anti-parallel by its very nature, and hence unsuitable for a modern CS curriculum. \cite{existential-type}}
\end{quote}

A problem within the US educational system is that, prior to entering the university, students have limited exposure to programming in general, and functional programming in particular. For an alarming number of US students, their first exposure to programming and computational thinking occurs after they graduate from high school. Yet in the world around them technology continues to advance and the complexity of software continues to increase. It should then perhaps come as no surprise that in recent years, US student interest in pursuing computer and information sciences has experienced an alarming decline.

\begin{quote}
\emph{Over the past five years[2001-2005], the percentage of ACT-tested students who said they were interested in majoring in computer and information science has dropped steadily from 4.5 percent to 2.9 percent.} \cite{ACT:2006-STEM-Pipeline}
\end{quote}

The 2012 ACT\footnote{ACT stands for American College Testing. The ACT is one of the primary organizations that tests high school students and assesses their readiness for college.} Profile Report indicates that planned education majors for students in the fields of computer science and mathematics (combined) is at 2 percent\footnote{The chart only displays whole numbers.}. Juxtaposed to this, the demand for computer science in the workforce continues to increase, and huge shortfalls are expected in the years to come.

Despite their syntactic simplicity and semantic elegance, there is a reluctance to teach functional programming as a first language, especially in the K-12 system\footnote{K-12 denotes Kindergarten through the $12^{th}$ grade.}. A primary reason for this is that ``interesting functional programs'' typically contain recursive definitions.

This design paper introduces \bricklayer, an API written in SML that provides a set of abstractions for creating LEGO artifacts which can be viewed using LEGO Digital Designer (LDD). The goal of \bricklayer\ is to create a problem space (i.e., a set of LEGO artifacts) that is accessible and engaging for programmers (especially novice programmers) while providing an authentic introduction to the functional programming language SML. The long term goal of \bricklayer\ is to provide a comprehensive and flexible set of modules that enable instruction to be tailored to the needs of a wide range of students. In particular, we hope to see \bricklayer\ used at all age levels where coding is taught (e.g., 8 years on up, including introductory university courses).

One key aspect of \bricklayer\ is that it provides a number of traversal functions. These traversals make it possible to build an interesting set of LEGO artifacts without having to confront the complexity of recursion! This feature enables \bricklayer\ to introduce functional programming to novices in a manner that is both gentle and engaging. This paper also describes an approach and philosophy to teaching functional programming and discusses how \bricklayer\ aligns with this approach and philosophy.

The remainder of the paper is as follows: Section \ref{section-related-work} discusses related work in developing environments and API's that facilitate teaching programming and computational thinking. Section \ref{section-teaching} argues for the use of functional languages to teach first-time programmers and describes key characteristics of computations suitable for programming assignments. Section \ref{section-bricklayer} gives an overview of the \bricklayer\ API and argues that the LEGO problem space is well-suited for introductory programming assignments. Sections \ref{section-predicates} through \ref{section-advanced-navigation} describe the various SML structures which make up \bricklayer, and Section \ref{section-conclusion} concludes.

\section{Related Work}\label{section-related-work}

In hopes of increasing interest in computer science, academia has developed languages and programming environments where skills in programming and computational thinking can be developed in ways that are fun and engaging to a broader student population. A fundamental issue confronted in the design of such languages and environments is that of \emph{authenticity}. Authenticity characterizes the degree to which a language realistically combines computational thinking with general-purpose programming. By this metric, syntax-free interfaces associated with languages such as Scratch \cite{Maloney:2010:Scratch} and Alice \cite{Dann:2006:ALICE} are considered to focus more on computational thinking. In an alternate approach, a general-purpose programming language such as Java, Jython\footnote{Jython is Java-based implementation of Python that has the ability to use Java classes.}, or SML can be extended with an API for the purposes of bringing a problem domain within reach of a certain computational thinking skillset. For example, CodeSpells \cite{Esper:2013:CodeSpells} teaches Java programming through first-person immersion in a 3D fantasy role-playing game. In CodeSpells, successful game play requires understanding and adaptation of spells, which are Java methods, belonging to a spellbook. Another example is the Media Computation API \cite{Guzdial:2003:Media-Computation-ITiCSE}, developed by Guzdial, that enables non-CS majors to construct Jython programs which manipulate photos in a variety of ways. The approach taken by \bricklayer\ is similar to that of the Media Computation API and brings a world of LEGO\textregistered\ artifacts within reach of authentic SML programming.

At Seton Hall University, a two course introduction to computer science, CS1 and CS2, has been developed in which programming assignments center around the creation of a Space Invader-like video game \cite{Morazan:Functional-Video-Games}. The courses follow an approach to teaching advocated by Felleisen, et al., and articulated in the text \emph{How to Design Programs}\cite{Felleisen:How-to-Design-Programs}, which places primary emphasis on design (not coding). Programs are written in dialects of Scheme, and DrScheme\footnote{DrScheme is now DrRacket} provides a rich programming environment in which programs can be developed and debugged. With the help of special libraries (called teachpacks) students begin developing their video game on the first day of class.

Picturing programs \cite{Bloch:Picturing-Programs} is another DrRacket-based framework for teaching introductory programming. The novel aspect of Picturing Programs is that, through a \emph{picturing-programs} teachpack, beginner-level Racket programs can be developed in which images themselves are treated as primitive values. Furthermore, images can be pulled in from any source including \url{http://images.google.com}. Basic operations on images include those that position images, flip images, rotate images, scale images, and overlay images. Additional functions provided include those that draw geometric shapes and crop images.

At Radboud University, a second year introductory course in functional programming is taught in which programming assignments revolve around implementing the behavior of a soccer team \cite{Achten:2011:SP:1972805.1972806}. The framework in which soccer games are played is called \emph{Soccer-Fun} which is implemented in the language Clean. A soccer game is a video game-like animation and is displayed in a GUI which is developed using the Clean Object I/O Library\cite{achten:CleanIO}. A central focus in Soccer-Fun is to implement the brain of a soccer player, a stateless function having eight parameters. Aside from its display capabilities, Soccer-Fun provides a domain specific language (DSL) consisting of a rich (and sophisticated) set of datatypes for modeling essential elements of soccer.  Students enrolling in the Soccer-Fun course have had training in C and Java and this provides them with sufficient maturity to understand the DSL of Soccer-Fun. In contrast, \bricklayer\ targets a much younger audience. \bricklayer\ assumes students have had no exposure to programming of any kind. \bricklayer\ also makes minimal assumptions about the mathematical background of students.

Yampa \cite{Courtney:Yampa} is a Domain-Specific Embedded Language (DSEL) for Functional Reactive Programming (FRP). It is implemented in Haskell as a combinator library and extends functional programming with the notion of \emph{time flow}. Yampa provides a first class component called a signal function which can be composed into networks using switching combinators. To facilitate the specification of larger signal networks, Yampa incorporates John Hughes' arrow framework. The combinator library that is Yampa provides a general framework for the implementation of continuous and discrete time systems. While its clarity, conciseness and modularity have been demonstrated via the implementation a Space Invader-like game it is difficult to imagine how Yampa, due to its sophistication, could be used in an introductory programming course.

\section{On Teaching First-time Programmers}\label{section-teaching}

As a programming language, SML's simplicity (when compared to languages like Java) makes it an attractive candidate for an introductory course on programming. SML's (stateless) core consists of (1) primitive types, values, and operators, (2) conditional expressions, (3) \emph{val} and \emph{fun} declarations, and (4) let-blocks. The key challenge faced in functional programming lies in understanding the dynamics of recursion. Though intellectually cleaner than invariant-based reasoning about loops, recursion is based on inductive reasoning to which students often have had insufficient prior exposure. For this reason, it is beneficial to introduce recursion in a gradual fashion only after sufficient dexterity has been attained in non-recursive programming.


A basic understanding of any programming language begins with an an introduction to primitive types, values, operations, and a handful of core programming constructs (e.g., expressions, declarations, conditions). These language constructs can be composed in a variety of ways to create simple programs (or program fragments) whose execution performs a single \emph{computational step}. For example, an expression that multiplies two numbers is considered to perform a single computational step. From such fundamentals one can begin the study of how to write a program whose execution will carry out a specific \emph{computation sequence} comprised of one or more computational steps.

Such computation sequences can be partitioned according to whether they describe \emph{fixed-length} or \emph{variable-length} computations. Fixed-length computations do not contain iterators and essentially perform a constant number of computational steps regardless of their input. For example, a function that converts Fahrenheit to Celsius can be implemented using a fixed-length computation. In contrast, the number of computational steps in variable-length computations can vary according to their input. For example, sorting a problem whose solution requires a variable-length computation. A key observation here is that in functional languages variable-length computations are typically achieved through recursion. In fact, one of the beautiful aspects of higher-order functions such as \emph{map} and \emph{fold} is that they relieve programmers of ``low level'' recursive details. In a similar spirit, \bricklayer\ provides primitives for placing LEGO bricks at specific locations as well as specifying general properties of cells within a virtual three dimensional space. LEGO artifacts can be built using such primitives that are large, complex and interesting, but nevertheless do not require students to define recursive functions.


\paragraph{Suitable Problems:} Aside from iteration, introductory programming also constrains, in a variety of ways, the kinds of programming problems that are \emph{suitable}. Suitable problem statements must be concise and may not make use of advanced data representations or programming language constructs. As a result, teachers often look to mathematical domains to provide the venue for studying both fixed-length as well as variable-length computation. Examples of suitable mathematical problems include: (1) do three integers form a Pythagorean triple, (2) conversions from base 10 to binary, octal or hexadecimal, (3) binary search in the form of 20 questions, (4) tests for primeness (e.g., Sieve of Eratosthenes), (5) perfect squares, (6) magic squares, and (7) termination experiments involving the Collatz conjecture. One concern is that such problems oftentimes do not sufficiently appeal to a broad student population. In contrast, the domain of LEGO artifacts accessible through \bricklayer, is more understandable and appealing to a larger audience. The following provides anecdotal evidence in support of this claim.

In the summer of 2014 a week-long workshop was held where 12 students, whose ages ranged from 9 to 14, were taught to program using \bricklayer\cite{2014:winter:techademy}. Two additional \bricklayer\ modules, \emph{Basic2D} and \emph{Basic3D}, were developed specifically for the target audience. Due to space constraints, the discussion of these modules lies beyond the scope of this paper. The interested reader is encouraged to visit the \bricklayer\ web site for more details\footnote{\url{http://faculty.ist.unomaha.edu/winter/Bricklayer/index.html}}. Feedback, in the form of a survey, revealed that students found the LEGO domain very interesting and a Bricklayer Coding Club has been created, at the request of the students and their parents, so these students as well as some younger students can continue to develop their programming and computational thinking skills.

It is interesting to note that in \bricklayer\ it is possible to build artifacts that have no physical LEGO counterpart (e.g., a rain effect can be created by suspending bricks in the air).  Our experience has been that students, especially younger students, initially build a physical LEGO artifact and then write code to create a corresponding artifact in virtual space. Thus, their thinking and approach to design is limited by physical reality. This connection (or lack thereof) is a central issue in high-consequence systems. Specifically, the evolution of a system design will oftentimes involve the replacement of hardware with software. A great concern is that software does not inherently conform to the laws of physics, and this can be the root cause of catastrophic failure. Appreciating this difference/fact is ultimately very important. In \bricklayer, beginning students initially program in a manner implicitly constrained by physical laws. For example, they limit the brick shapes they use to the standard shapes, they do not leverage overwriting in their implementations, and so on. It is interesting to observe the ``ah-ha'' moment when they realize that in the virtual world they are not bound by the laws of the physical world.


\paragraph{Problems, Examples, and Study:} The construction of implementations that solve problems is a key form of feedback employed in programming classes. When viewed along this dimension, a course can be seen as a sequence of \emph{programming assignments} (aka problems) separated by \emph{study} (the development of understanding and computational thinking skills needed to solve a targeted class of problems).

At the heart of study lie germane and compelling \emph{examples} highlighting certain forms of computational thinking. In this context we define an \emph{example} to be a triple consisting of (1) a problem statement, (2) a problem solution, and (3) an instantiated articulation of various forms of computational thinking that can be used to reach the problem solution from the problem statement. A \emph{problem} then can be seen as simply the prefix (i.e., the problem statement) of an example.

In order to be \emph{effective}, a example-problem sequence should have the following attributes:

\begin{enumerate}
 \item examples and problems are suitably coupled,
 \item problems cover a targeted set of learning goals,
 \item the change in difficulty between two related elements in the sequence falls within an appropriate tolerance, being neither too difficult nor too easy, and
 \item problems and examples are engaging to students.
\end{enumerate}

To achieve this, one needs a domain that is \emph{example rich} and \emph{problem dense}. This enables the construction of effective example-problem sequences that are both accurate and precise with respect to a given set of learning objectives. In Section \ref{section-bricklayer}, we argue that the LEGO domain accessible through \bricklayer\ is \emph{example rich} and \emph{problem dense}.

\section{\bricklayer}\label{section-bricklayer}
\bricklayer\ \cite{Bricklayer} is an API, written in SML, that is being developed at the University of Nebraska at Omaha. \bricklayer\ provides a basic set of abstractions for creating LEGO\textregistered\ artifacts which can be viewed using LEGO Digital Designer (LDD). The rational for \bricklayer\ is premised on the assumption that a LEGO mindset is conducive to the study of programming for the following reasons:

\begin{enumerate}
\item \emph{engaging} -- Generally speaking, LEGO has a universal appeal. People like LEGOs -- both students and their parents.
\item \emph{determined} -- Many students have had prior exposure to the construction of physical LEGO artifacts. For example, it is not uncommon for elementary school students to have assembled LEGO artifacts consisting of hundreds (even thousands) of pieces. Thus, the association of extensive instruction sequences with LEGO has already been established. This association fosters patience and endurance with respect to LEGO-related activities. Such a mindset is beneficial to (\bricklayer) programming.

\item \emph{concrete} -- A LEGO artifact inhabits a discrete space having a physical manifestation. This provides an environment where students can, in many cases, ``play with'' and develop a prototype-based physical understanding of a problem before expressing its solution in code.

\end{enumerate}

In the sections that follow, we hope to convince the reader that the artifacts that can be created using \bricklayer\ belong to a domain that is \emph{example rich} and \emph{problem dense} (as discussed in Section \ref{section-teaching}). Users interact with \bricklayer\ through four SML structures (aka modules) named: (I) Predicate, (II) BrickFunction, (III) BasicNavigation, and (IV) AdvancedNavigation. These structures have been ordered according to their computational sophistication and expressive power. A summary of the knowledge of SML that is prerequisite for the \bricklayer\ structures is shown in Table \ref{table-bricklayer}.

Note that recursion is first introduced in the BasicNavigation structure (III). In Table \ref{table-bricklayer}, we designate a function declaration to be \emph{simple-recursive} if it implements the behavior of a loop. In this paper, we will refer to such recursive functions as \emph{simple recursion}.

\begin{figure*}[htb!]
\centering
\subfloat[Predicate\label{first}]{
\includegraphics[trim = 0mm 20mm 5mm 0mm, clip, scale=0.14]{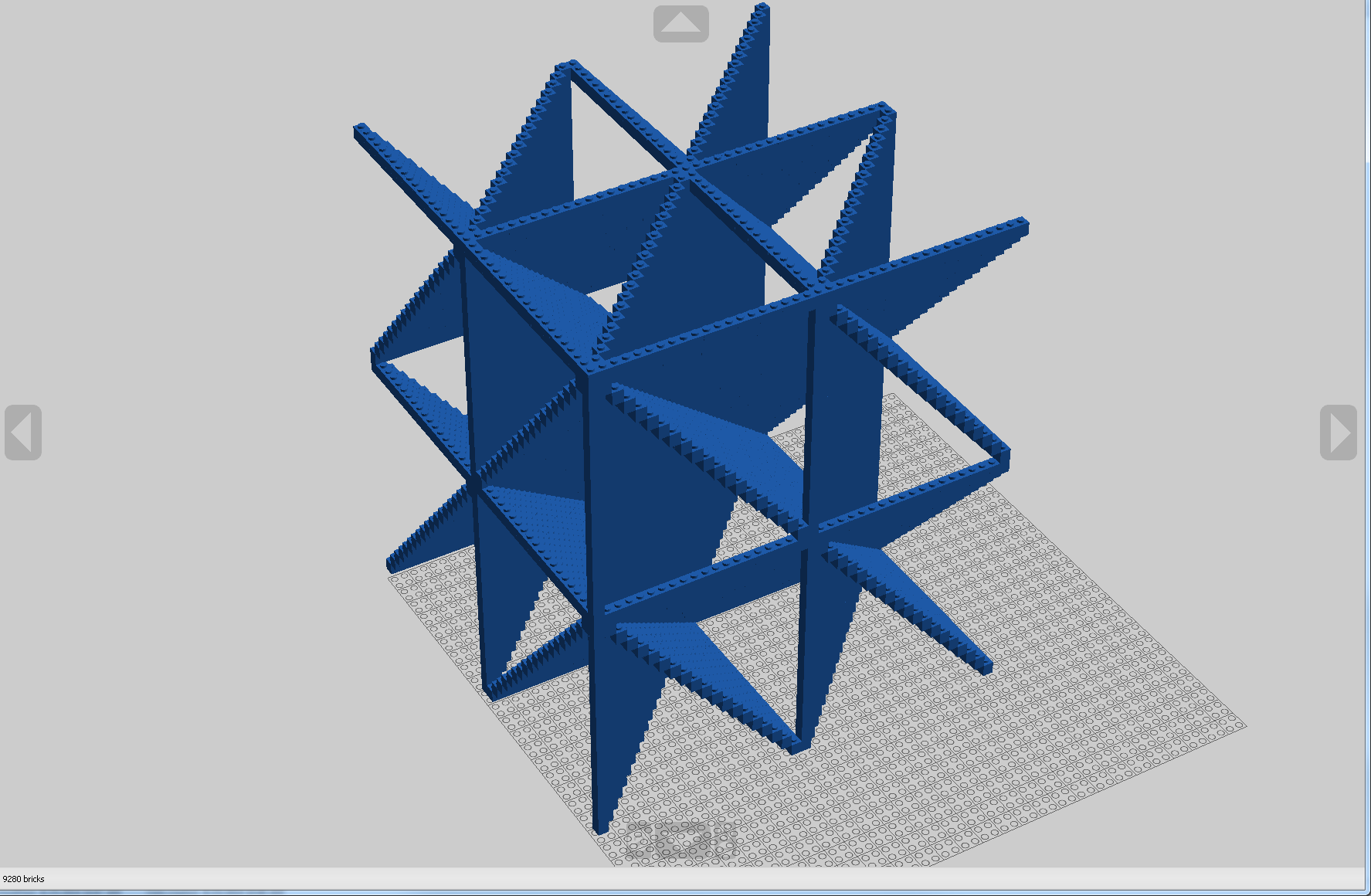}
}
\subfloat[BrickFunction\label{second}]{
\includegraphics[trim = 0mm 20mm 5mm 0mm, clip, scale=0.14]{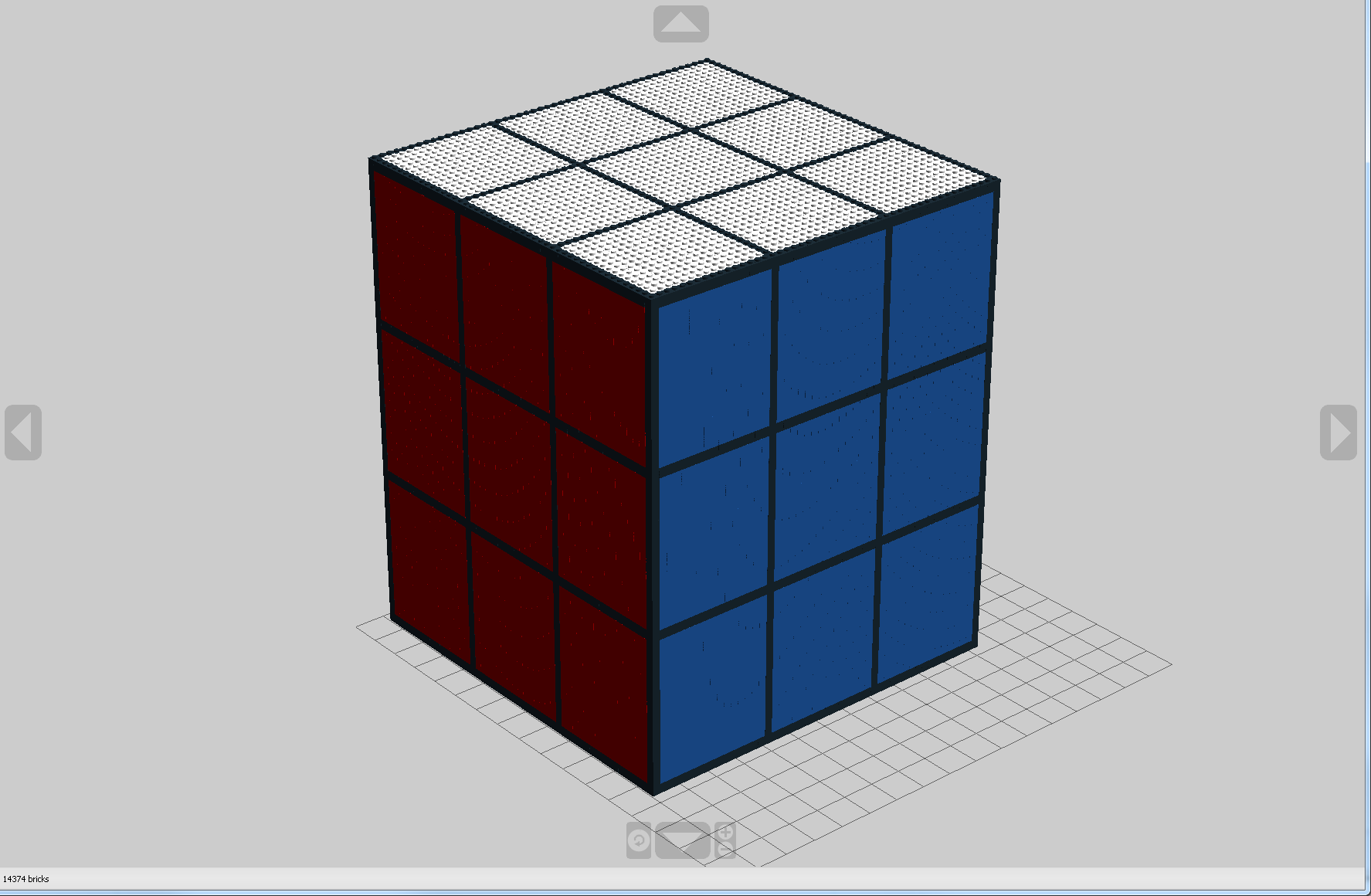}
}

\subfloat[BasicNavigation\label{third}]{
\includegraphics[trim = 0mm 20mm 5mm 0mm, clip, scale=0.14]{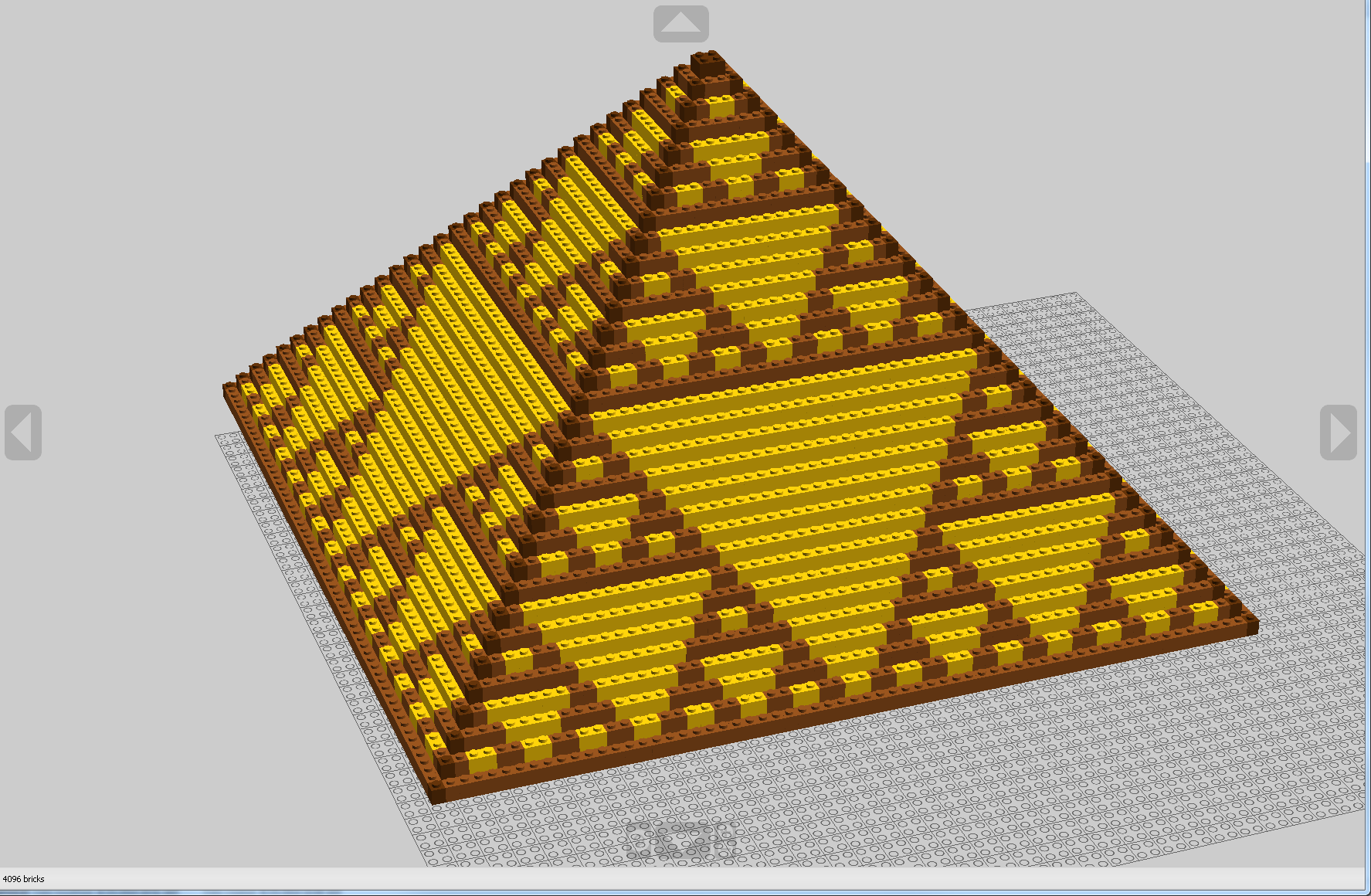}
}
\subfloat[AdvancedNavigation\label{fourth}]{
\includegraphics[trim = 0mm 20mm 5mm 0mm, clip, scale=0.14]{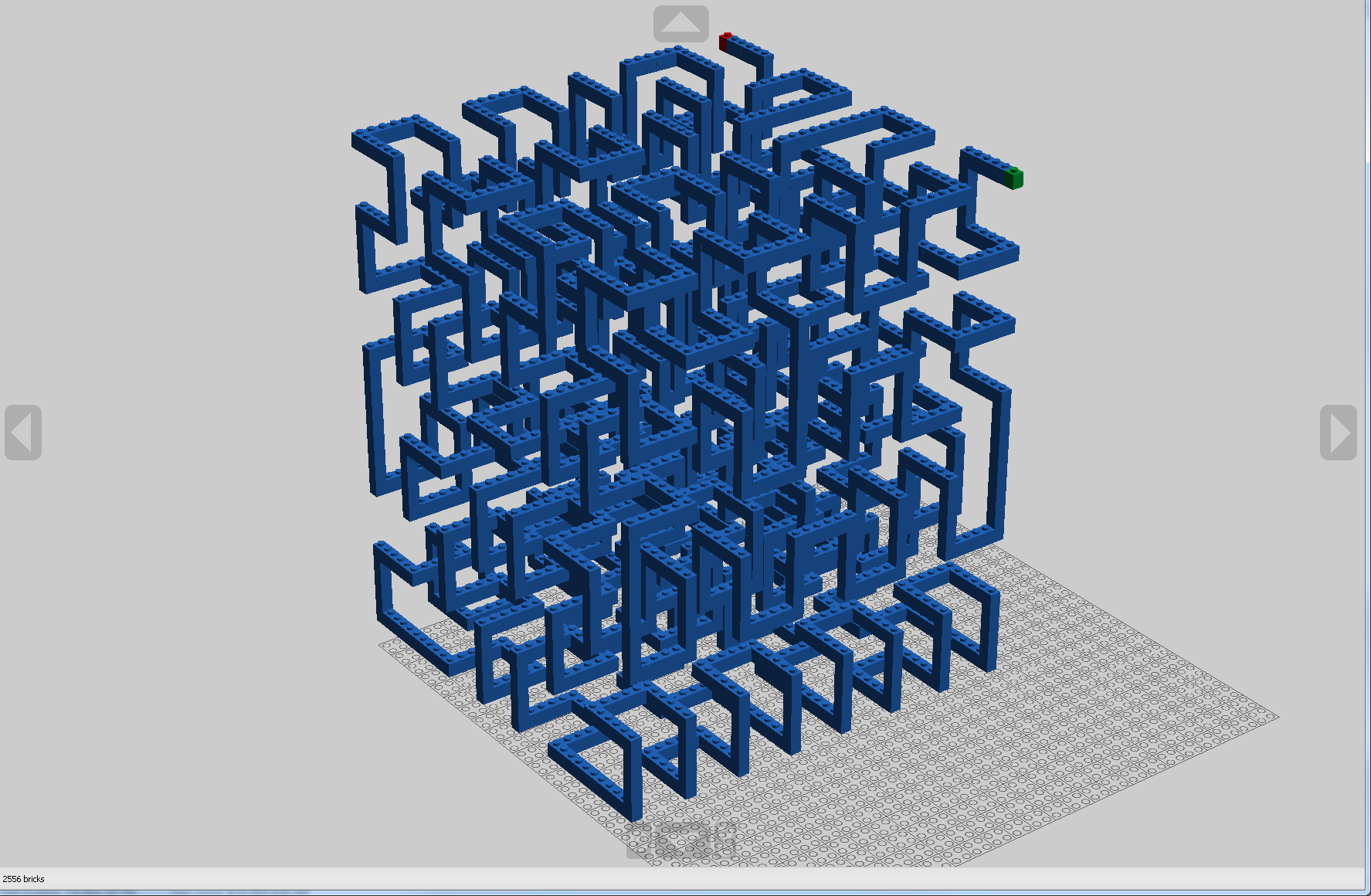}
}
\caption{Artifacts created using Bricklayer and displayed using LDD}\label{fig-bricklayer-structures}
\end{figure*}

Examples of the kinds of structures that can be created using the \bricklayer\ structures and displayed using LDD is shown in Figure \ref{fig-bricklayer-structures}.

\begin{table*}[htb!]
\centering
\begin{tabular}{@{}lcccc@{}}\toprule
                                 & \textbf{I} & \textbf{II} & \textbf{III} & \textbf{IV} \\ \midrule
\multicolumn{5}{l}{\textbf{primitive types}} \\ \midrule
\qquad integer                   & $\surd$    & $\surd$     & $\surd$      & $\surd$     \\ \hline
\qquad Boolean                   & $\surd$    & $\surd$     & $\surd$      & $\surd$     \\ \hline
\qquad string                    &            &             & $\surd$      & $\surd$     \\ \hline
\multicolumn{5}{l}{} \\ \midrule
\ \textbf{variables}                        & $\surd$    &  $\surd$    & $\surd$      & $\surd$     \\ \midrule
\multicolumn{5}{l}{} \\
\multicolumn{5}{l}{\textbf{operators}} \\ \midrule
\qquad arithmetic                & $\surd$    &  $\surd$    & $\surd$      & $\surd$     \\ \hline
\qquad comparison                & $\surd$    &  $\surd$    & $\surd$      & $\surd$     \\ \hline
\qquad logical                   & $\surd$    &  $\surd$    & $\surd$      & $\surd$     \\ \hline
\multicolumn{5}{l}{} \\
\multicolumn{5}{l}{\textbf{expressions}} \\ \midrule
\qquad arithmetic                & $\surd$    &  $\surd$    & $\surd$      & $\surd$     \\ \hline
\qquad logical                   & $\surd$    &  $\surd$    & $\surd$      & $\surd$     \\ \hline
\qquad let-blocks                & $\surd$    &  $\surd$    & $\surd$      & $\surd$     \\ \hline
\qquad conditional expressions   &            &  $\surd$    & $\surd$      & $\surd$     \\ \hline
\qquad sequence                  &            &             & $\surd$      & $\surd$     \\ \hline
\qquad output (i.e., print)      &            &             & $\surd$      & $\surd$     \\ \hline
\multicolumn{5}{l}{} \\
\multicolumn{5}{l}{\textbf{declarations}}   \\ \midrule
\qquad val-dec                   & $\surd$    & $\surd$     & $\surd$      & $\surd$     \\ \hline
\qquad non-recursive fun-dec     & $\surd$    & $\surd$     & $\surd$      & $\surd$     \\ \hline
\qquad simple-recursive fun-dec  &            &             & $\surd$      & $\surd$     \\ \hline
\qquad recursive fun-dec         &            &             &              & $\surd$     \\
\bottomrule
\end{tabular}
\caption{A partial overview of the intersection between SML concepts and \bricklayer\ modules.}\label{table-bricklayer}
\end{table*}

\subsection{Separating Computational Concerns}

\bricklayer\ provides some very simple ways to specify \emph{virtual spaces}, which are 3D spaces whose integer coordinates range from $(0,0,0)$ to $(x,y,z)$ where $x,y,z > 0$. In addition, \bricklayer\ provides a number of functions for interacting with virtual spaces. Most notably, \bricklayer\ provides a variety of simple functions for \emph{(generically) traversing} a virtual space (user-defined traversals are also possible).

As an aside, it is worth noting that the importance of generic traversal and related computational ideas have been recognized in a variety of areas. In the field of strategic programming, generic traversal enables the separation of conditional rewrite rules from the algorithmic details surrounding their application. In declarative programming and rewriting systems the technical details surrounding search and rule application are handled by the computational model underlying the language. \bricklayer\ introduces such \emph{separation of computational concerns} into the SML core. Through \bricklayer\ traversals it is possible to build a wide variety of three dimensional LEGO artifacts without having to confront the complexity of recursion.

\section{Predicates}\label{section-predicates}

The signature of the Predicate structure is given in Appendix \ref{appendix-predicate}. The most important function exported by the Predicate structure is: \emph{show}. This function when called with

\begin{enumerate}
\item an integer $n$ denoting the length of the side of a cube,
\item a function (i.e., a predicate) from $(x,y,z)$ coordinates to Boolean values, and
\item a brick value
\end{enumerate}

\noindent will
\begin{enumerate}
\item create a virtual cube within Bricklayer whose coordinates range over $\mbox{(0,0,0),\dots,(n-1,n-1,n-1)}$,
\item traverse the virtual cube's coordinate space and apply the given function to each coordinate, placing the given brick value at every coordinate for which the predicate evaluates to true,
\item output the resulting virtual structure to an lxfml format recognized by LDD, and
\item perform a system call loading LDD with the lxfml file created.
\end{enumerate}

\textbf{Example:} When executed, the following SML code will create and display a $20\times20\times20$ cube consisting of blue 1-bit LEGO bricks.

\medskip
\begin{minipage}[b]{3in}
\begin{lstlisting}
fun cube(x,y,z) = true;

Predicate.show(20,cube,Pieces.BLUE);
\end{lstlisting}
\end{minipage}
\hspace*{5mm}
\begin{minipage}{2in}
\includegraphics[trim = 0mm 20mm 5mm 0mm, clip, scale=0.14]{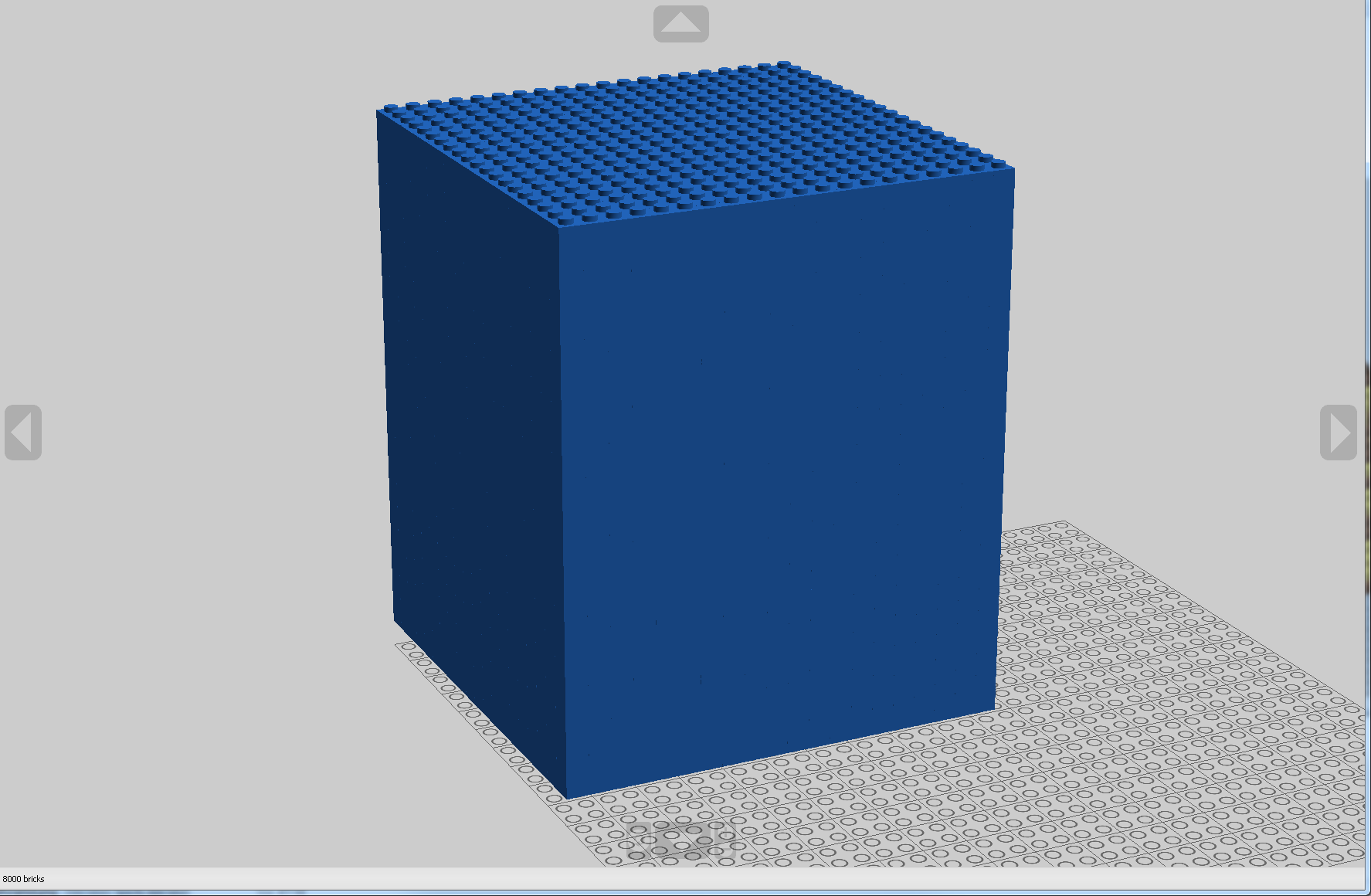}
\end{minipage}

\bigskip
\textbf{Example:} When executed, the following SML code will create and display a $20\times20$ square in the X-Z plane consisting of red 1-bit LEGO bricks.

\medskip
\begin{minipage}[b]{3in}
\begin{lstlisting}
fun square(x,y,z) = y = 0;

Predicate.show(20,square,Pieces.RED);
\end{lstlisting}
\end{minipage}
\hspace*{5mm}
\begin{minipage}{2in}
\includegraphics[trim = 0mm 20mm 5mm 0mm, clip, scale=0.14]{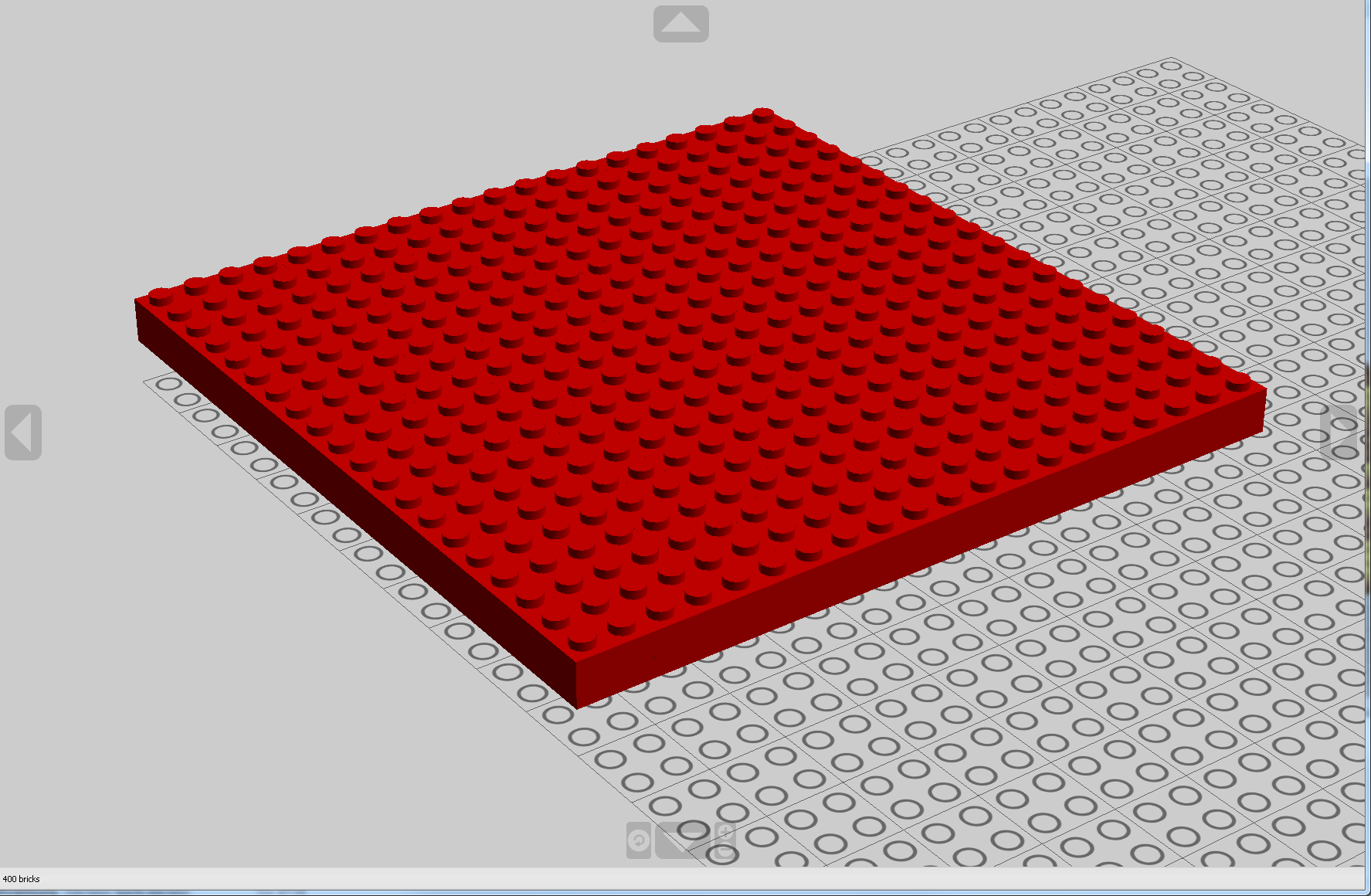}
\end{minipage}

\medskip

From a technical standpoint, the set of computations that can be studied using the Predicate structure center on expressions involving Boolean and comparison operators. Additional programming constructs, such as helper functions and local variables, may be used to structure code and increase readability. This set of computations, which is a subset of the set of \emph{fixed-width computations}, is immense and has incredible expressive power. From a theoretical perspective, this is to be expected.

\subsection{Providing Feedback through \bricklayer\ Comparison Functions}

\begin{figure*}[htb!]
\centering
\subfloat[Correct artifact = $A$\label{correct}]{
\includegraphics[trim = 0mm 20mm 5mm 0mm, clip, scale=0.14]{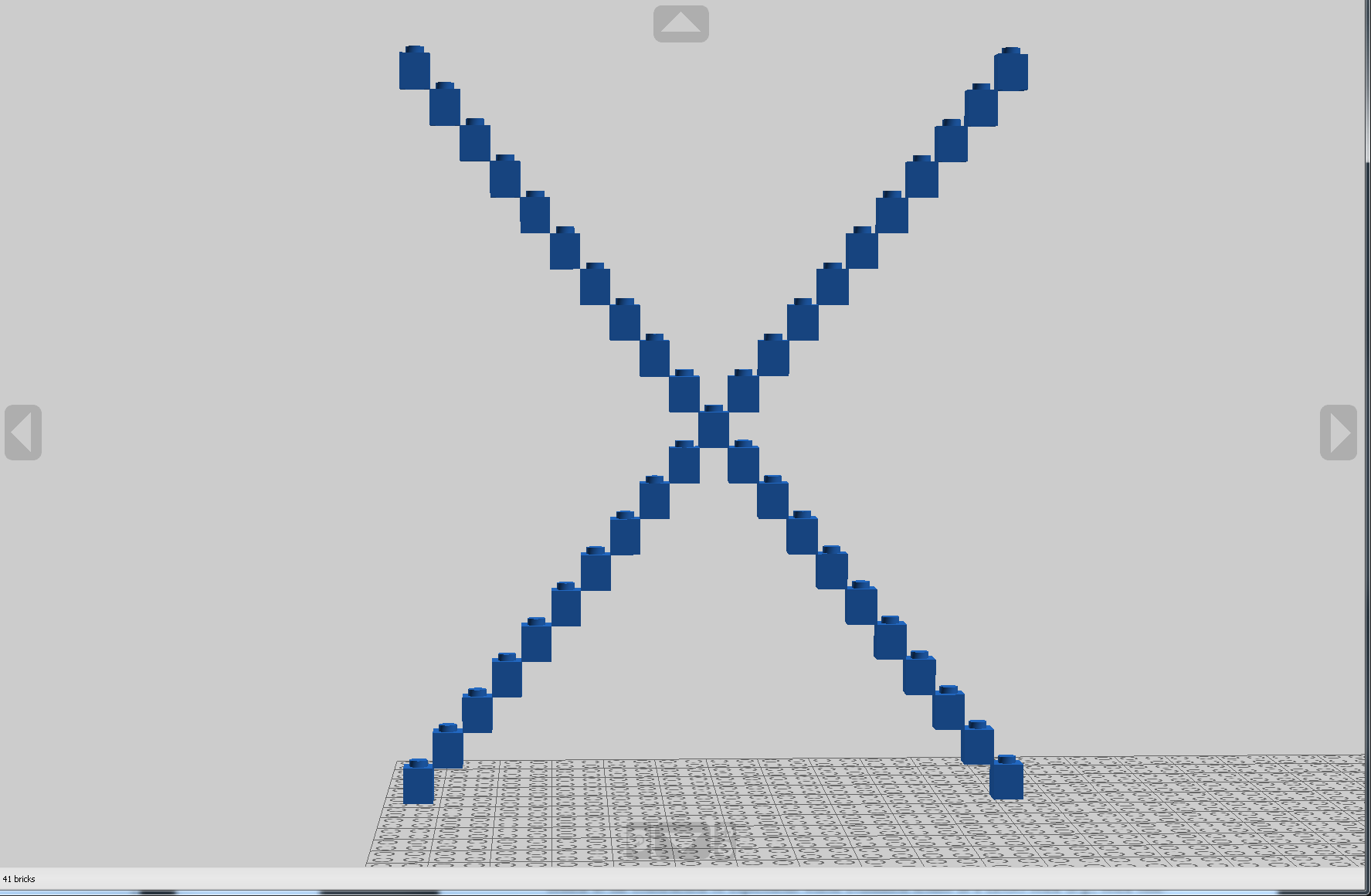}
}
\subfloat[Incorrect artifact = $B$\label{incorrect}]{
\includegraphics[trim = 0mm 20mm 5mm 0mm, clip, scale=0.14]{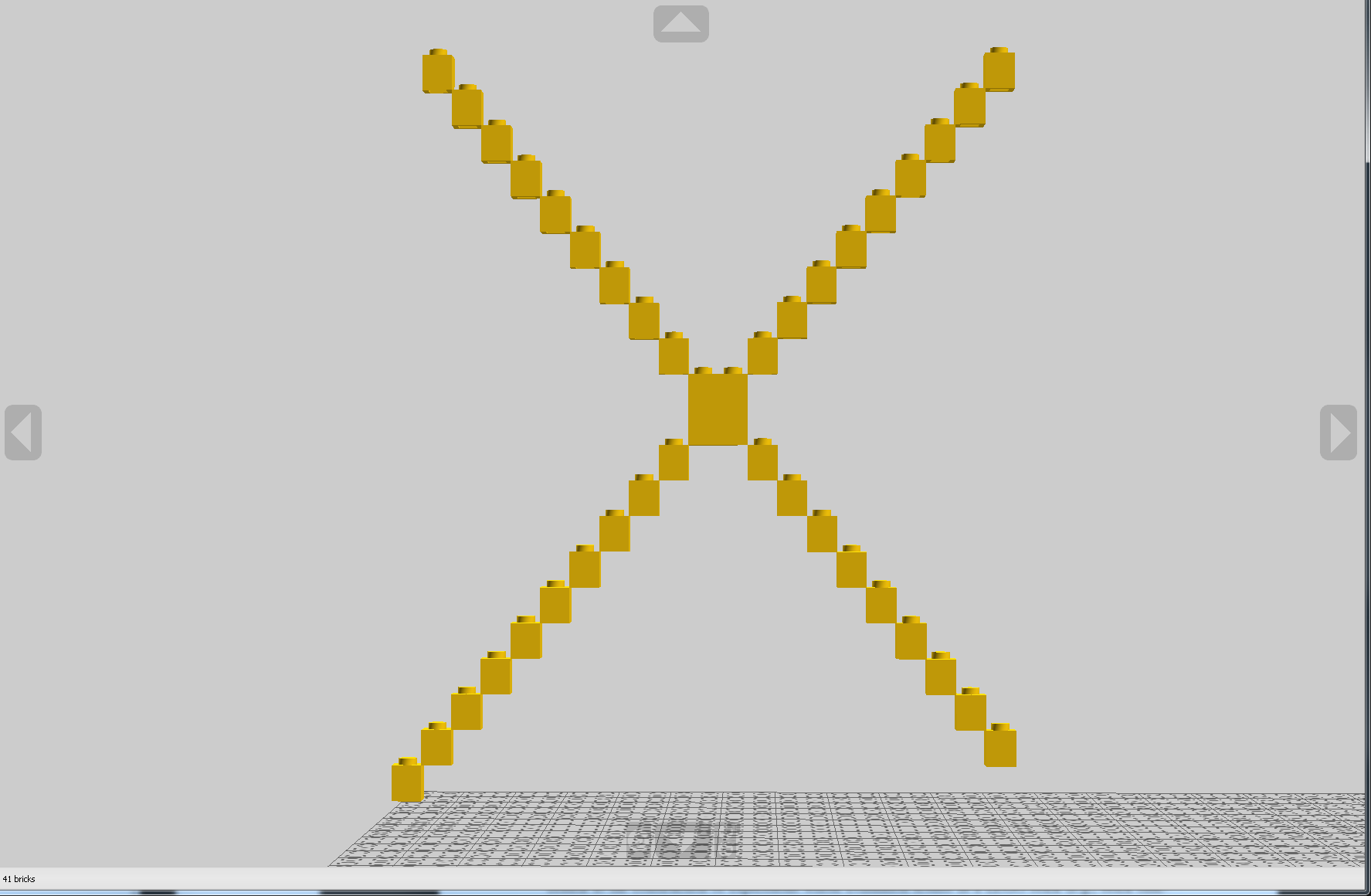}
}

\subfloat[$A \oplus B$ \label{xor}]{
\includegraphics[trim = 0mm 20mm 5mm 0mm, clip, scale=0.14]{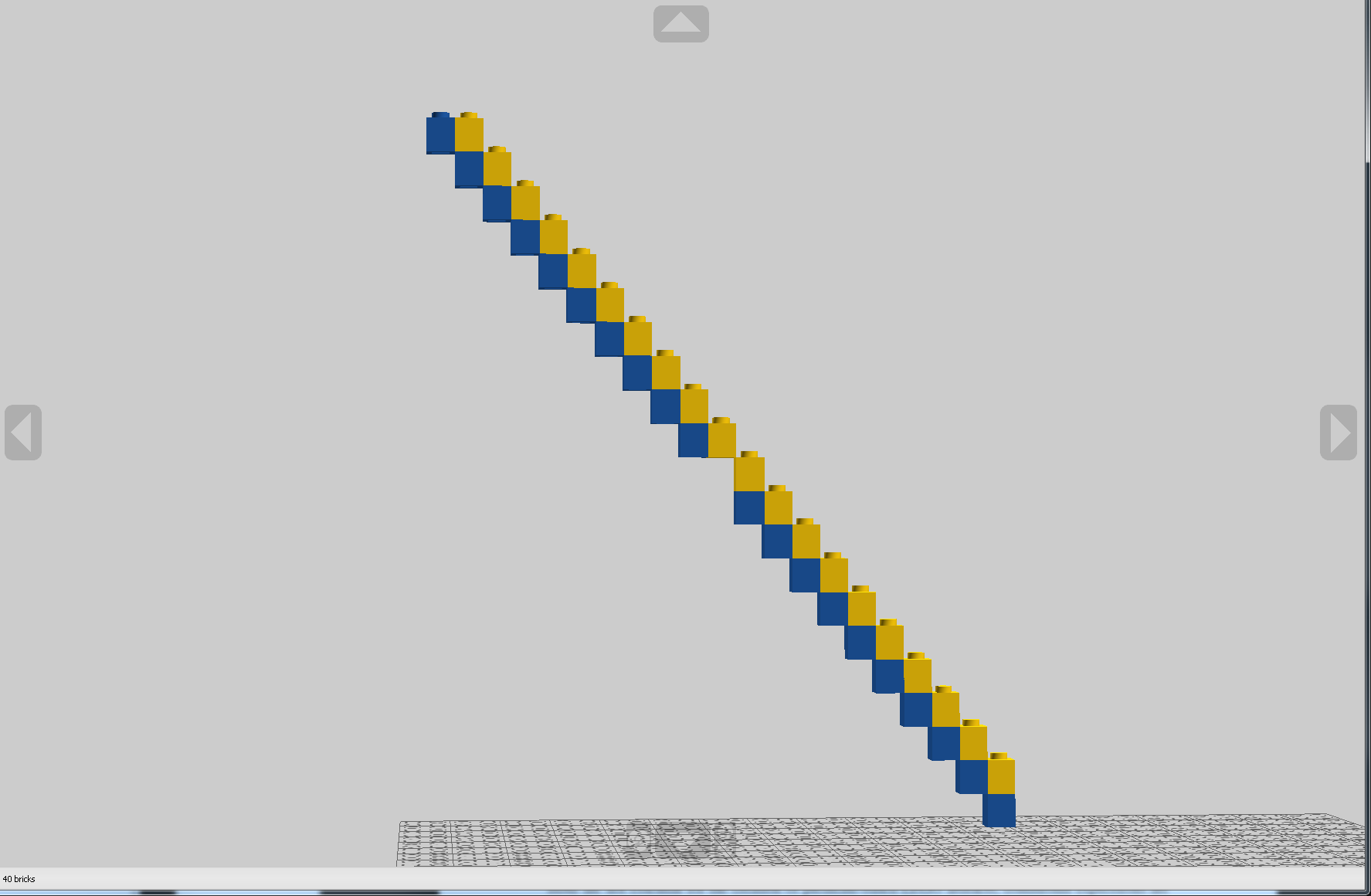}
}
\subfloat[$A \cup B$\label{union}]{
\includegraphics[trim = 0mm 20mm 5mm 0mm, clip, scale=0.14]{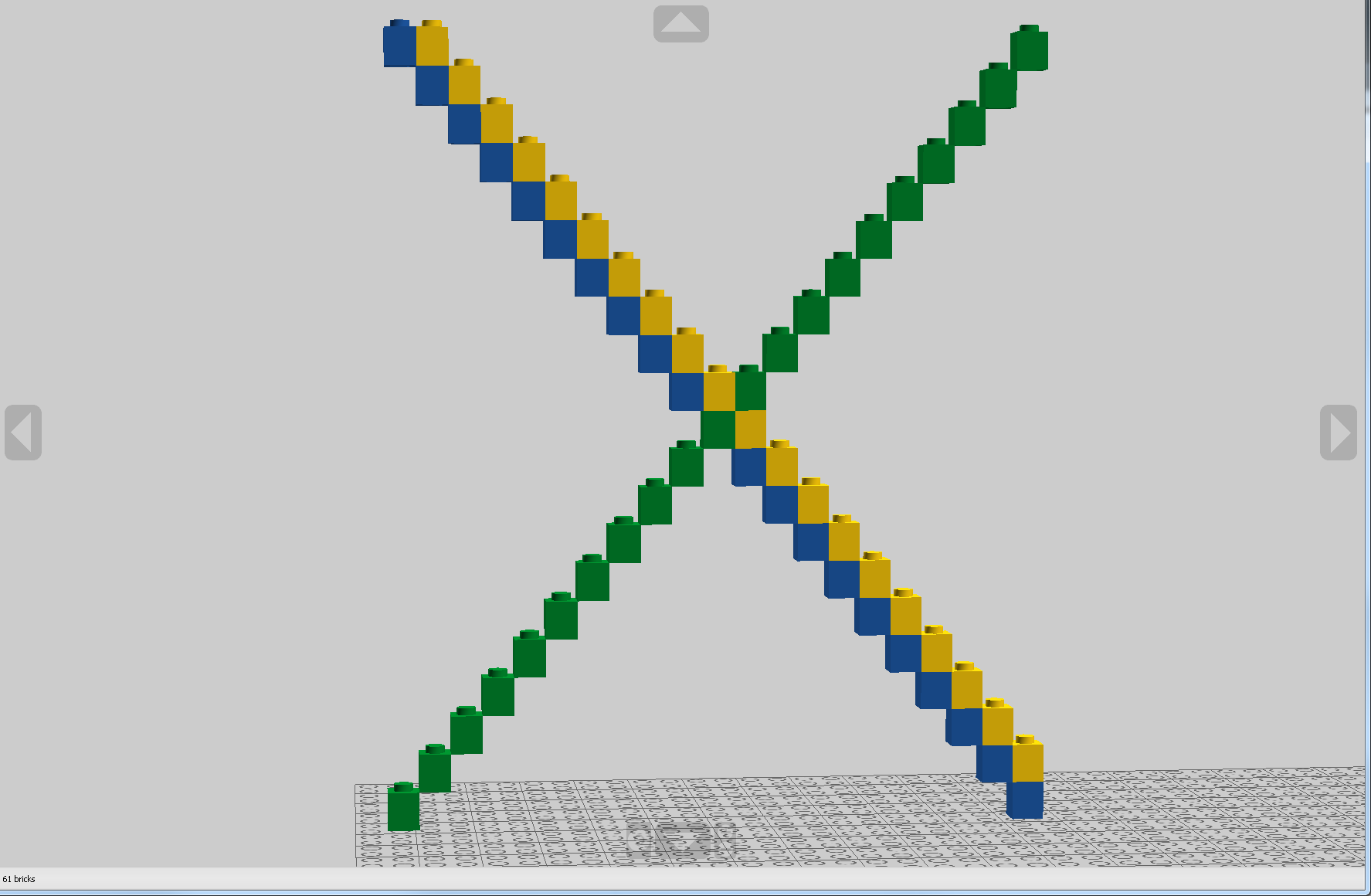}
}

\subfloat[$A \cap B$\label{intersection}]{
\includegraphics[trim = 0mm 20mm 5mm 0mm, clip, scale=0.14]{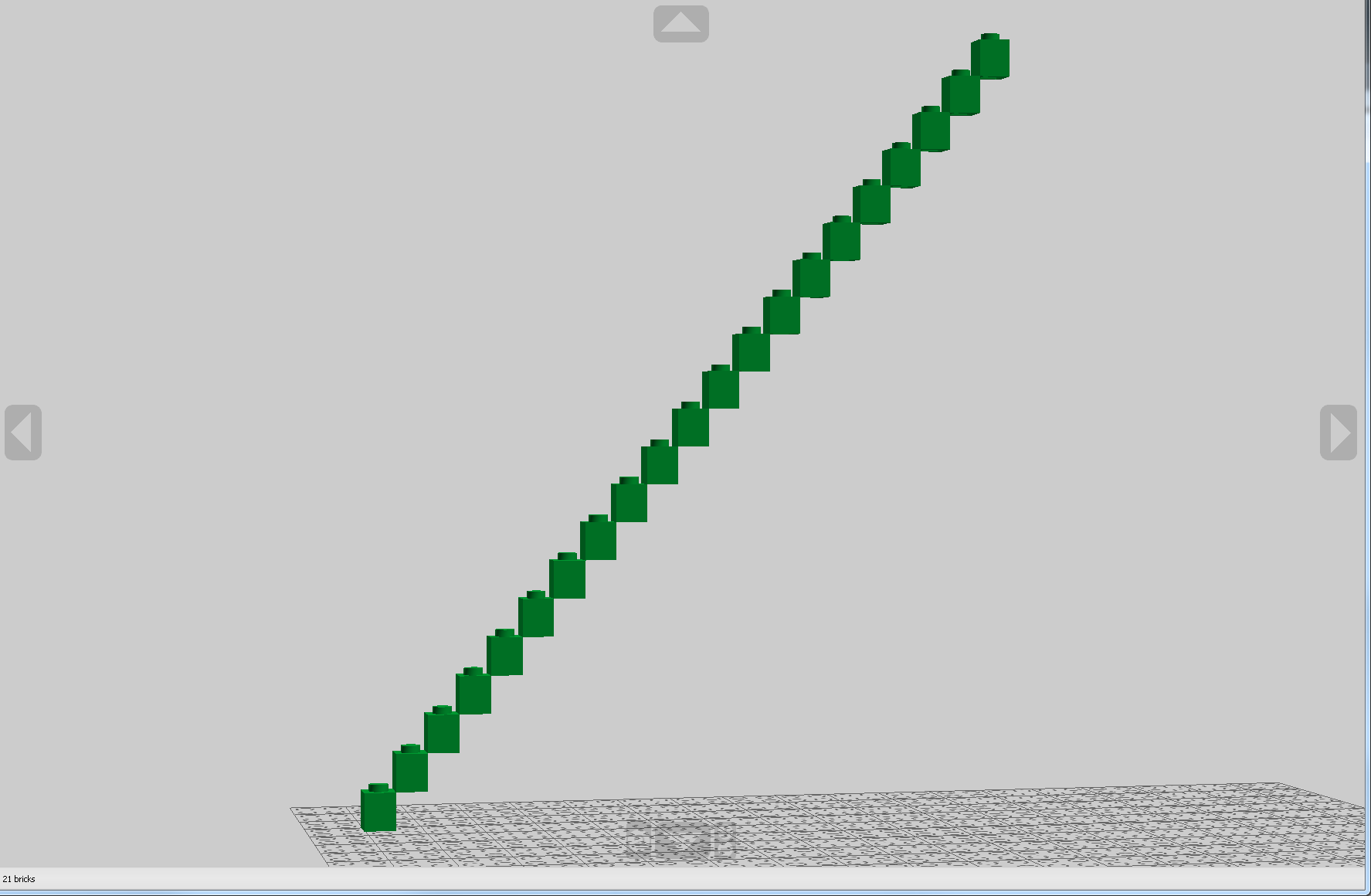}
}
\subfloat[$A - B$\label{difference}]{
\includegraphics[trim = 0mm 20mm 5mm 0mm, clip, scale=0.14]{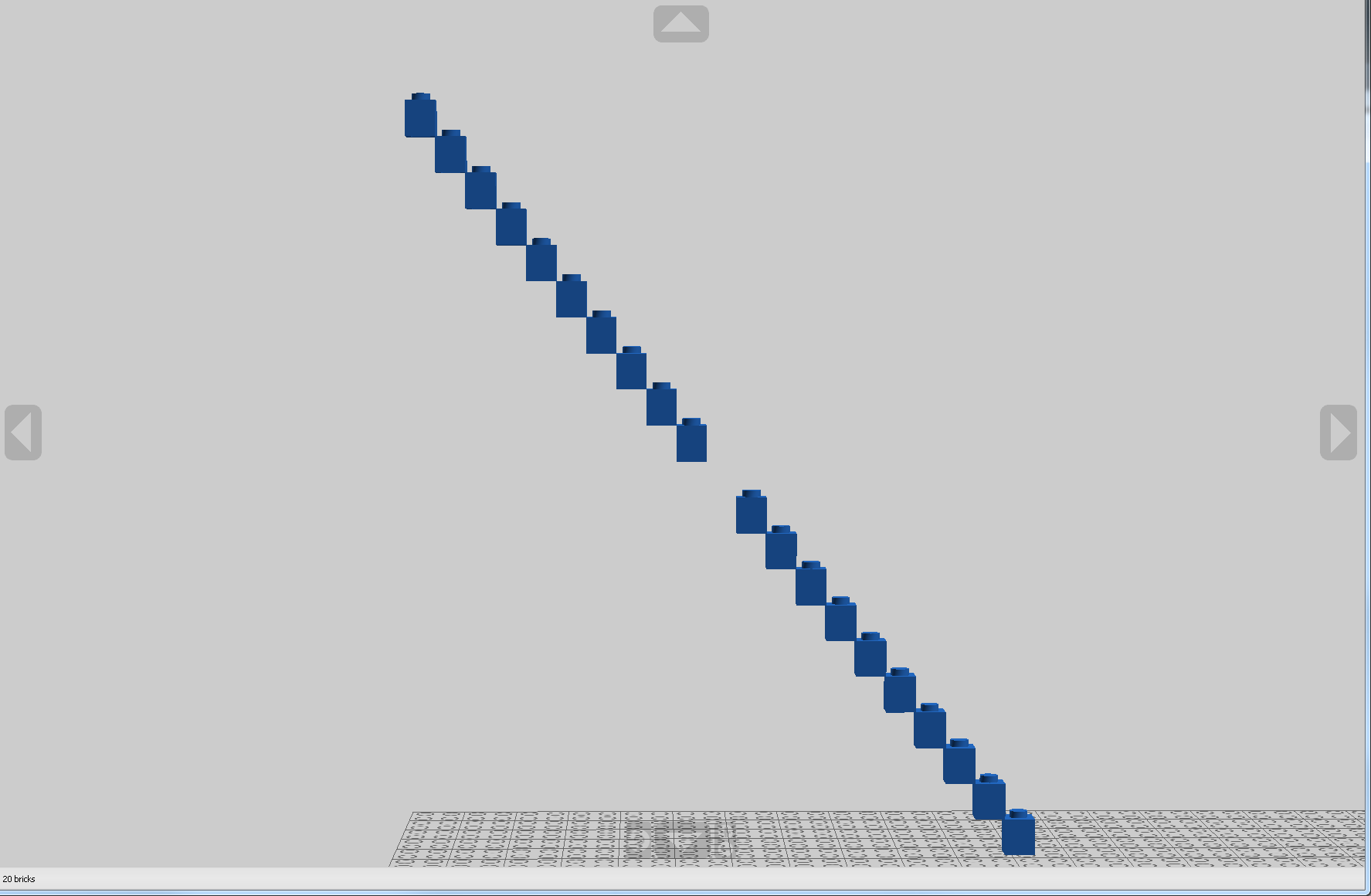}
}
\caption{Comparing and contrasting student solutions with instructor solutions.}\label{fig-predicate-comparisons}
\end{figure*}

In addition to \emph{show}, the Predicate structure exports the functions \emph{xor}, \emph{union}, \emph{intersection},  and \emph{difference}. These functions enable users to examine differences between two LEGO artifacts (e.g., an artifact in a problem description and the artifact produced by their code). In particular, a student can use these functions to compare (and contrast) their solution to a problem with the instructor's solution. Note, that predicates create monochromatic artifacts. Thus, without loss of generality distinct colors (i.e., bricks) can be assigned to predicates. For example, the solution artifact can be composed of blue bricks, the student artifact can be composed of yellow bricks, and cells that are occupied in both artifacts can be mapped to green bricks. It is this convention that is used in the discussion that follows.

Figure \ref{correct} shows the LEGO artifact that is the solution to a programming assignment. Figure \ref{incorrect} shows the LEGO artifact created by the student's program. In this case, the student's solution contains an off-by-one error. In practice, such errors are common and their discovery without the use of \bricklayer\ comparison functions can involve detailed counting (e.g., counting the number of LEGO bricks in the base of a triangle). Figures \ref{xor} through \ref{difference} show the various LEGO artifacts resulting from comparing the correct artifact with the student's artifact. Note that the error shown shifts the entire top-left to bottom-right diagonal one position to the right and therefore the result of the set difference operation shown in Figure \ref{difference} is not empty. The nature of the error is also confirmed by the union operation shown in Figure \ref{union}.

\bricklayer\ comparison functions provide users with control over the bricks used to create the resulting artifact. In Figure \ref{fig-predicate-comparisons}, (1) blue bricks are used to denote bricks belonging to the instructor solution, (2) yellow bricks are used to denote bricks belonging to the student's solution, and (3) green bricks are used to denote bricks that the instructor's solution and the student's solution have in common.

\subsection{Relationships between Triangles and Loops}
\begin{figure}[htb!]
\centering
\subfloat[$x >= y$\label{fig-first-triangle}]{
\includegraphics[trim = 0mm 20mm 5mm 0mm, clip, scale=0.11]{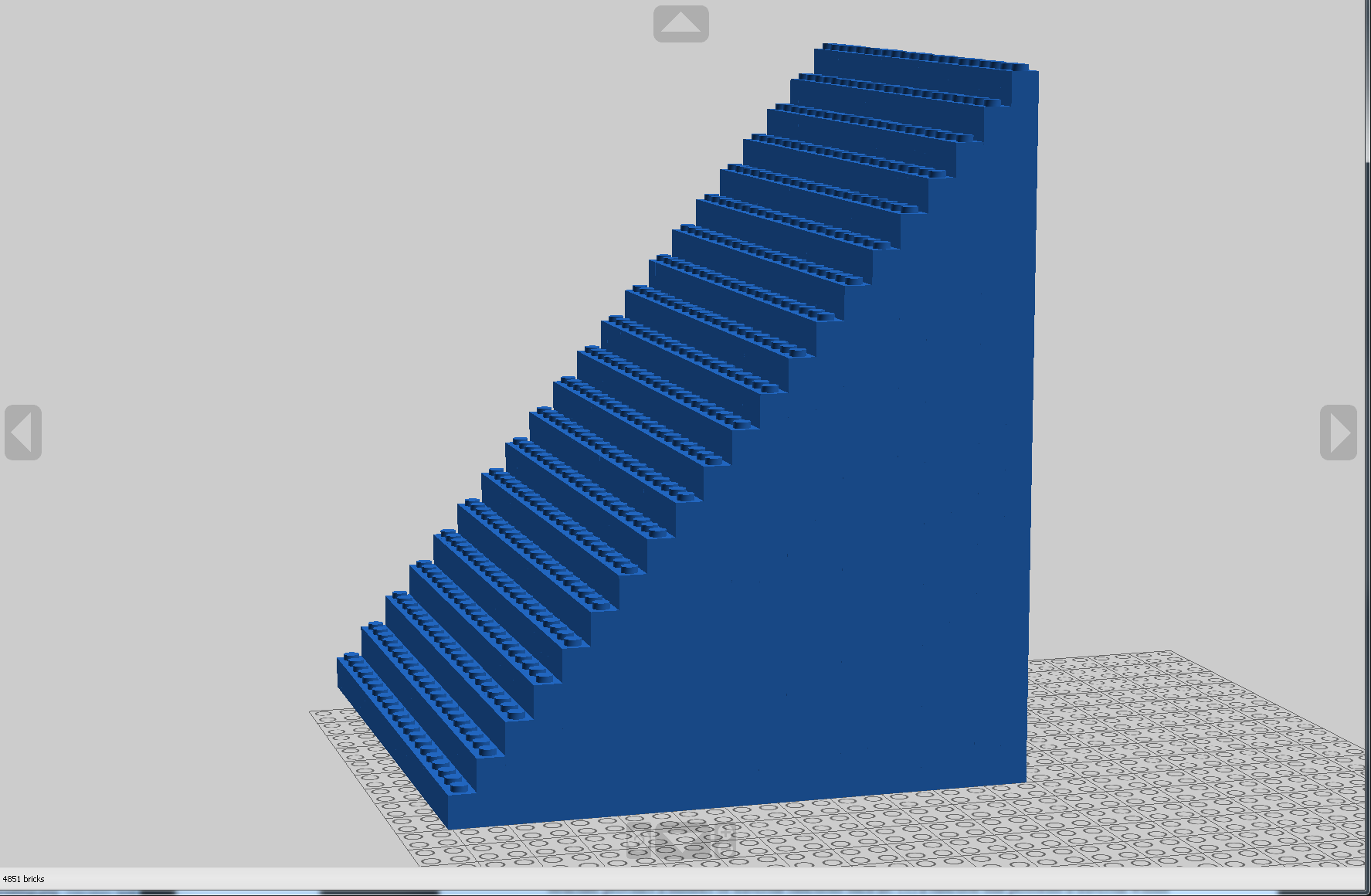}
}
\subfloat[$x+y < 21$\label{fig-second-triangle}]{
\includegraphics[trim = 0mm 20mm 5mm 0mm, clip, scale=0.11]{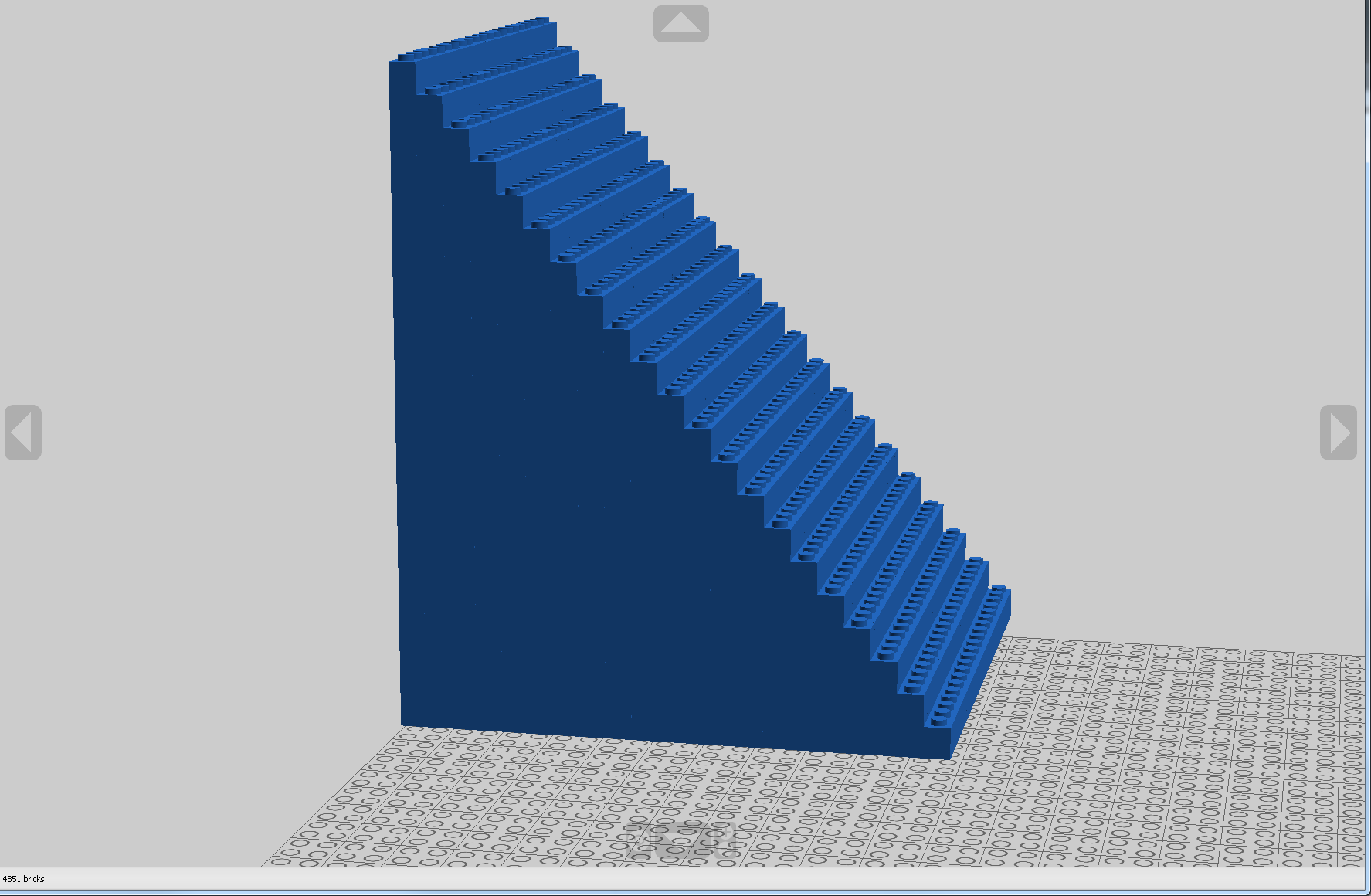}
}
\subfloat[$a \cap b$]{
\includegraphics[trim = 0mm 20mm 50mm 0mm, clip, scale=0.11]{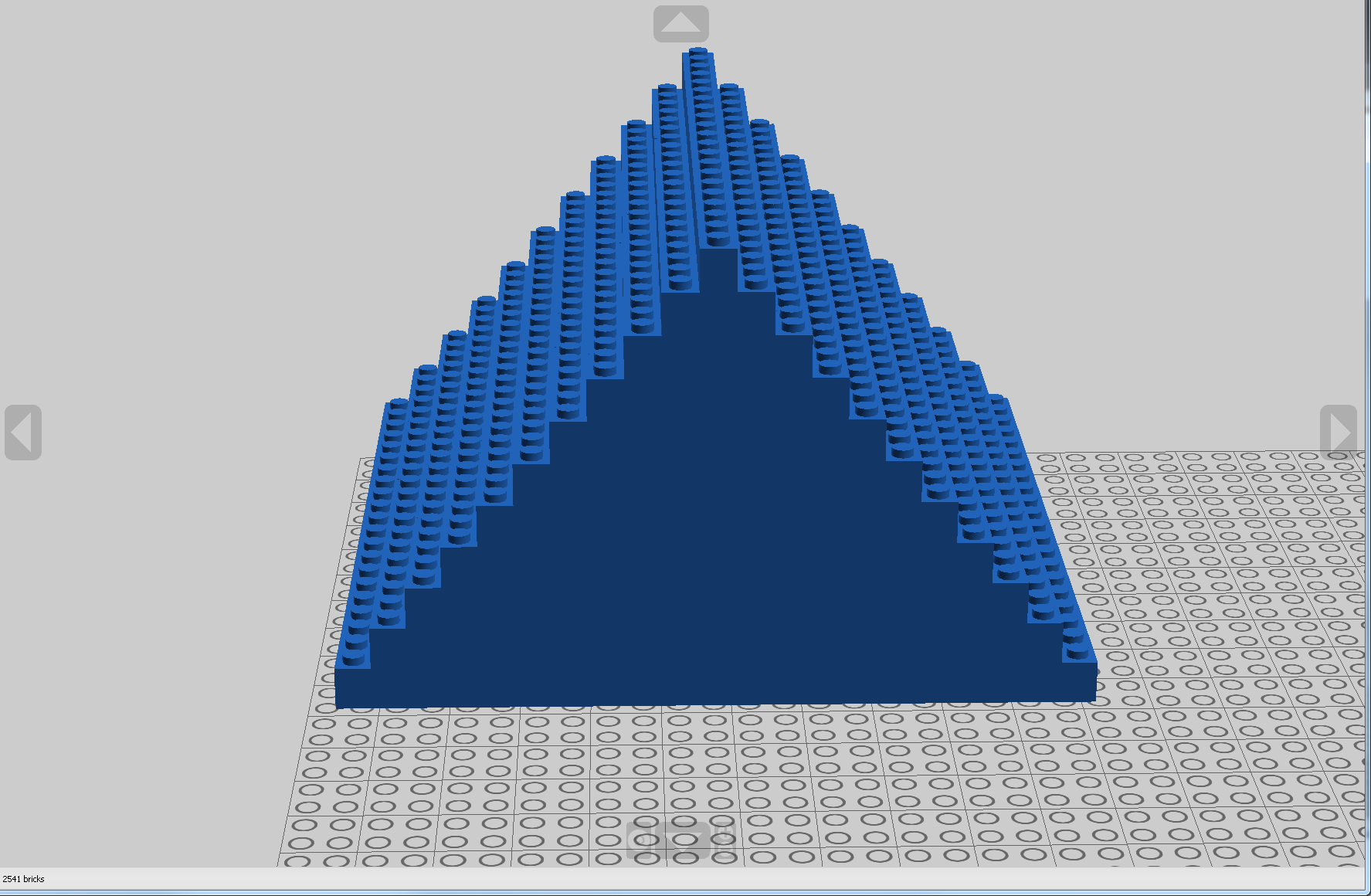}
}
\caption{Virtual space: $(0,0,0)\dots(20,20,20)$}\label{fig-simplePyramid}
\end{figure}

The construction of triangles and pyramids, in their various orientations and forms involve computational thinking similar to that underlying nested for-loops. Consider for example, the predicate $p(x,y,z) = x + y < 21$ whose structure is shown in Figure \ref{fig-second-triangle}. The computational thinking underlying this construction shares similarities to the computational thinking used to create imperative loops of the form:

\begin{center}
\begin{tabular}{l}
for i = 0 to 20 \\
\qquad  for j = 0 to 20 - i
\end{tabular}
\end{center}

The looping pattern shown above can be used to implement the bubblesort algorithm. Furthermore, counting the number of bricks placed in the triangle in Figure \ref{fig-second-triangle} requires computational thinking similar to that which is required to perform a big-$\mathcal{O}$ analysis of the running time of algorithms such as bubblesort.

\section{Brick Functions}\label{section-brick-functions}
The BrickFunction structure exports only a single function called \emph{show}. The behavior of \emph{BrickFunction.show} is similar to \emph{Predicate.show}. The only difference being that \emph{BrickFunction.show} accepts only two inputs: (1) an integer $n$ denoting the side of a cube, and (2) a total function $f$ from $(x,y,z)$ coordinates to brick values.

A limitation of the \emph{Predicate.show} is that objects are created using a single type of LEGO brick (e.g., blue, green, etc.). \emph{BrickFunction.show} removes this limitation by processing functions whose return values are LEGO bricks (instead of simple Boolean values). As a result, LEGO artifacts can now be created using multiple brick types (i.e., different colors and shapes).

When using the BrickFunction structure students encounter a compelling problem domain requiring the incorporation of conditional expressions into their programming skill sets.
While \emph{conditional expressions} are not essential for the creation of predicate-based LEGO artifacts, conditional expressions are central to the construction of expressions whose evaluation results in a LEGO brick (e.g., brick functions). Conditional expressions provide users with control over what kind of brick is placed in a given cell in the virtual space. However, with this ability comes the responsibility for designating which cells are to remain empty. Bricklayer provides an empty brick for this purpose.

\subsection{Periodic Functions and Equivalence Classes}
\begin{figure}[htb!]
\centering
\includegraphics[trim = 0mm 20mm 5mm 0mm, clip, scale=0.30]{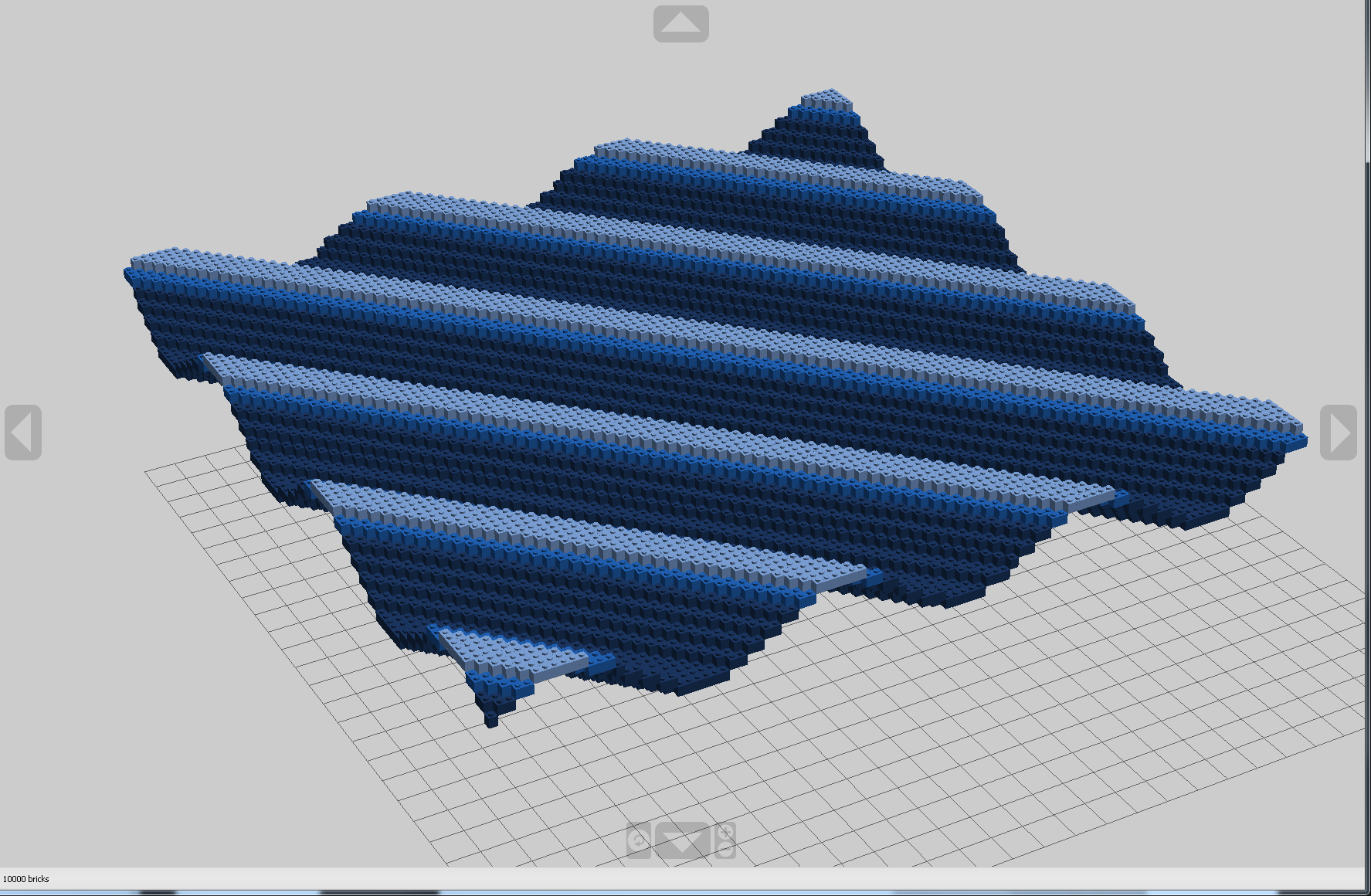}
\caption{Sine waves with shading.}\label{fig-diagonalSineWaves}
\end{figure}

Periodic functions (such as modulus and sine) can be used to create repeating patterns within LEGO artifacts. Periodic functions can be used in all \bricklayer\ levels. For example, the artifact in Figure \ref{first} results from a simple predicate that uses the \emph{mod} operator. It should be noted however that the effects of periodic functions are more compelling when dealing with multi-colored artifacts. The Rubik's cube\footnote{Special thanks to Jared Knust and Colin Kramer for providing this example.} shown in Figure \ref{second} is created by composing the \emph{mod} operator with conditional coloring. The artifact in Figure \ref{fig-diagonalSineWaves} is created using the sine wave function and mapping different heights of the curve to various shades of blue.

\begin{figure*}[htb!]
\begin{lstlisting}
fun bigCheckerboard boardSize squareSize =
  let
    (* assumes there exists an integer c such that boardSize = c * squareSize *)

    fun equivClass v = (v div squareSize) mod 2

    fun brickFunction(x,y,z) =
      let
        val black = equivClass x = equivClass z
      in
        if y = 0 then
           if black then Pieces.BLACK
           else Pieces.ORANGE
        else Pieces.EMPTY
      end
  in
    BrickFunction.show(boardSize, brickFunction)
  end;
\end{lstlisting}

\begin{center}
\includegraphics[trim = 0mm 20mm 5mm 0mm, clip, scale=0.3]{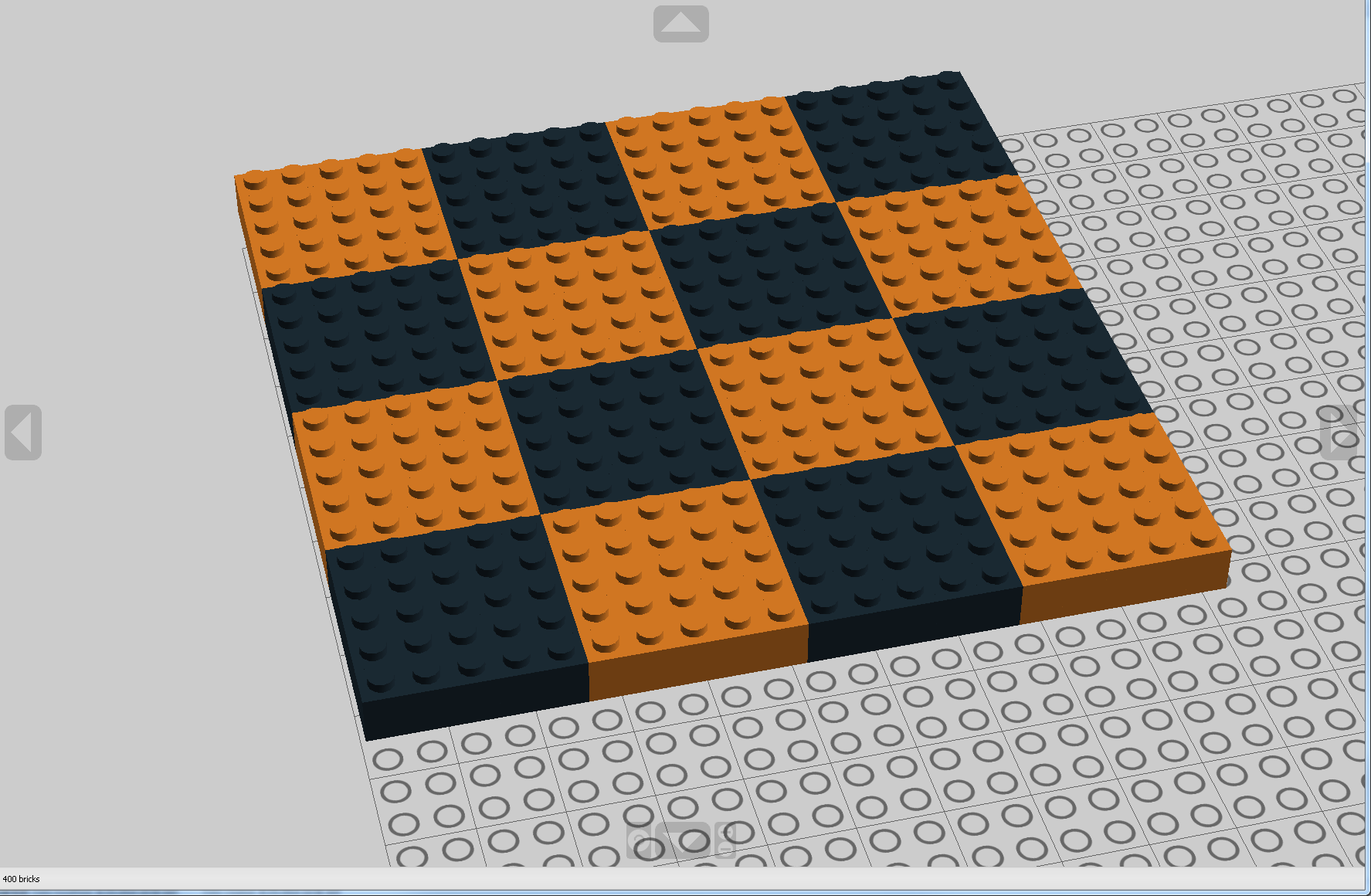}
\end{center}

\caption{Using equivalence classes to create a checkerboard having big squares.}\label{fig-bigCheckerboard}
\end{figure*}

Operations such as \emph{div} as well as ``within range'' comparisons can be used to group coordinates into equivalence classes. Mappings can then be created from equivalence classes to colors. Figure \ref{fig-bigCheckerboard} shows how \emph{div} and \emph{mod} operations can be combined to create a checkerboard where each square of the board consists of 25 LEGO bricks. The example contains a nested conditional expression where the outer conditional serves to define the ``domain of discourse'' for the inner conditional.

\section{Basic Navigation}\label{section-basic-navigation}
It is in the BasicNavigation structure where simple recursion is first introduced. The BasicNavigation structure, whose signature is given in Appendix \ref{appendix-predicate}, provides a number of traversal functions such as: (1) a function that performs a traversal within a given bounding box, and (2) a set of functions that traverse a two dimensional space for a fixed value of the third dimension. In addition, \emph{access} and \emph{update} functions are provided for cell-level interaction with the virtual space. A function is also provided that draws a ``smooth'' line between two given points. It is through these functions that students are gradually given control over how a virtual space is traversed and how cells visited during a traversal are populated. In the end, the objective is to give students complete control over the order in which cells are visited during a traversal. For example, simple recursion similar to nested for-loops can be used to iterate over the cells in a virtual space.

\begin{figure}[htb!]
\centering
\subfloat[Nested Cubes\label{fig-nested-cubes}]{
\includegraphics[trim = 0mm 20mm 5mm 0mm, clip, scale=0.17]{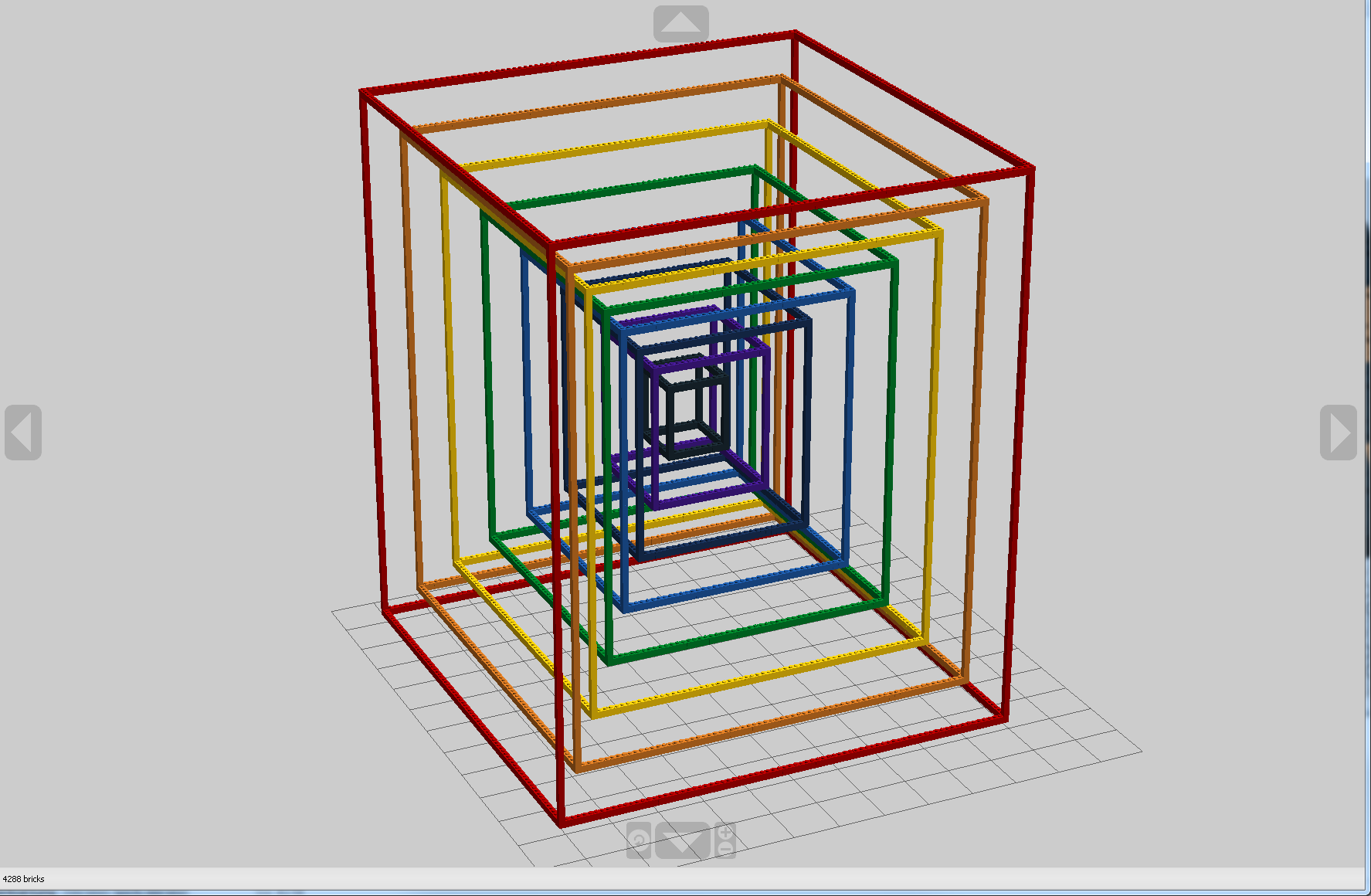}
}
\subfloat[Season's Greetings\label{fig-seasons-greetings}]{
\includegraphics[trim = 0mm 20mm 5mm 0mm, clip, scale=0.17]{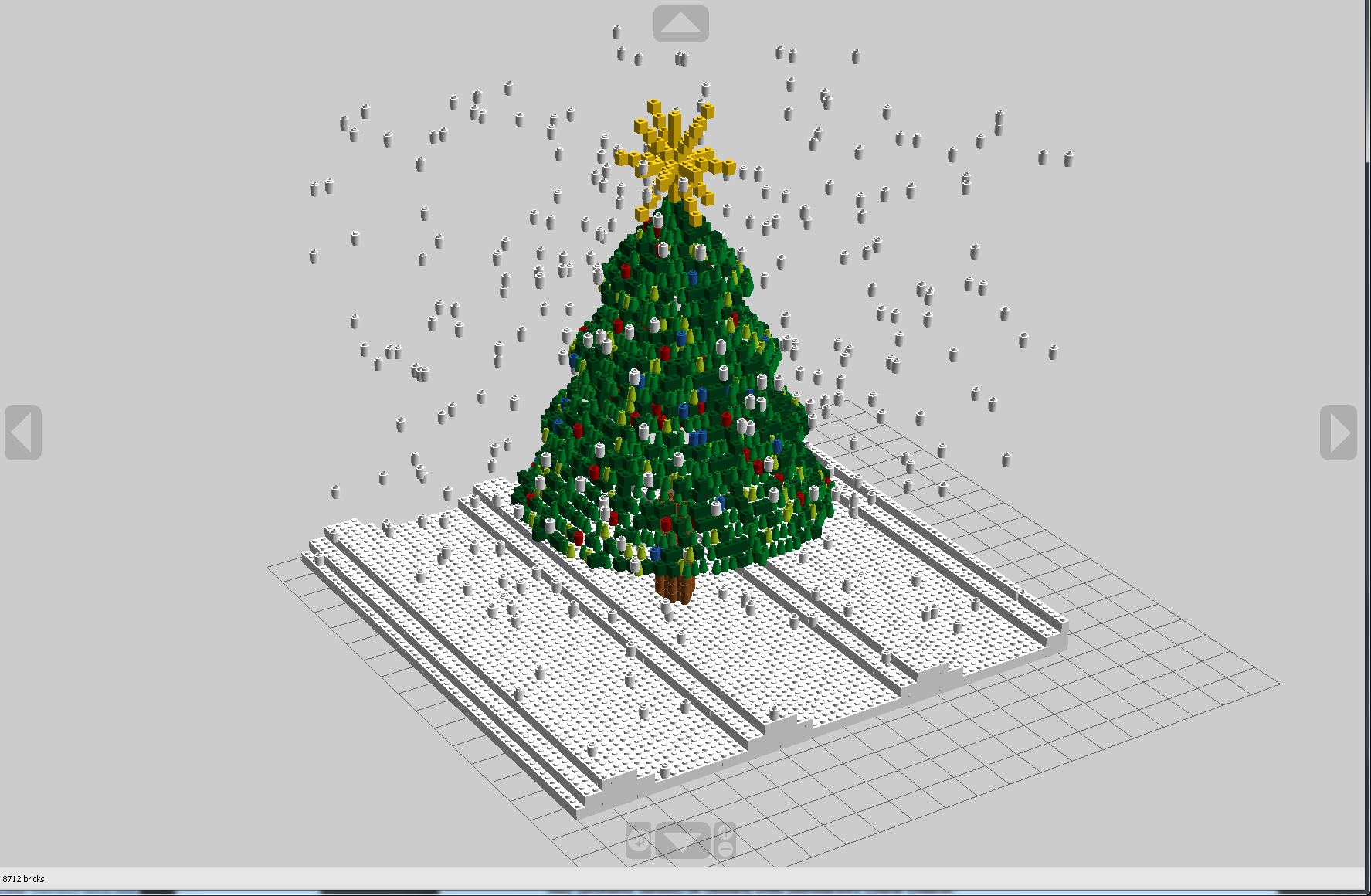}
}
\caption{Applications of simple-recursive functions.}\label{fig-simple-recursive}
\end{figure}

Figure \ref{fig-nested-cubes} shows an artifact that can be created using \emph{simple recursion} (see Section \ref{section-bricklayer}) that shrinks the bounding box (e.g., $(lo,lo,lo)$ $\dots$ $(hi,hi,hi)$) within which a wire-frame cube is to be created. Figure \ref{fig-seasons-greetings} is a composite artifact in which simple recursion is used to create the green portion of the Christmas tree which consists of a sequence of increasingly smaller green cones. The remaining portions of the artifact are produced in a non-recursive fashion (e.g., a random function is used to generate falling snow).

\begin{figure}[htb!]
\centering
\includegraphics[trim = 0mm 20mm 5mm 0mm, clip, scale=0.25]{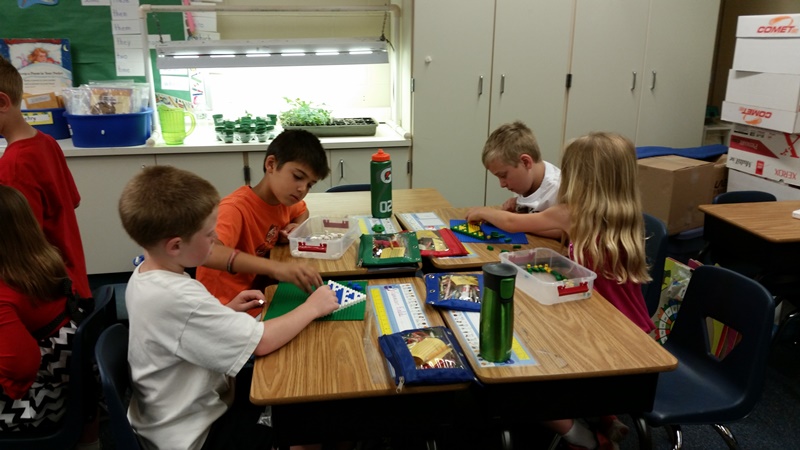}
\includegraphics[trim = 0mm 20mm 5mm 0mm, clip, scale=0.25]{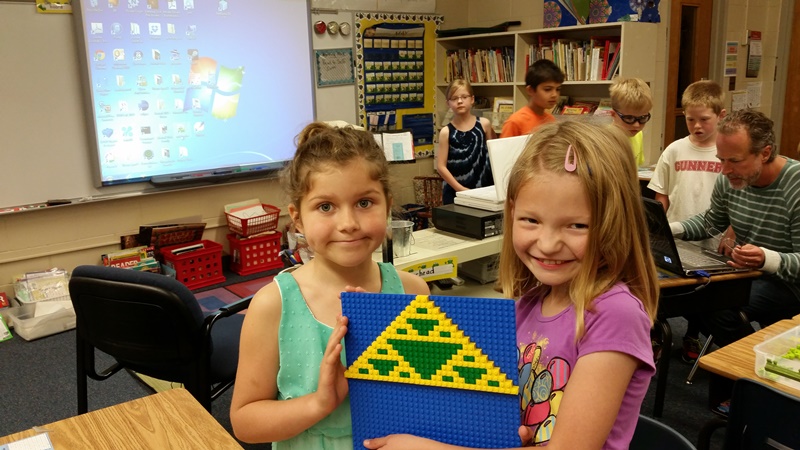}

\caption{Second grade (8-9 year old) students constructing the Sierpinski Gasket}\label{fig-sierpinski-lego}
\end{figure}

The access function provided by the BasicNavigation structure enables the contents of cells in the virtual space to be inspected. This opens the door to computations where the order in which cells are visited (during a traversal) is important. The artifact in Figure \ref{third}, which is based on the Sierpinski gasket, is an example of a computation in which order of traversal is important and must be explicitly controlled by the student code. There were only two simple recursive functions used in the construction of this artifact. There are several algorithms that can be used for creating a Sierpinski gasket. One possibility is to implement the gasket as a Lindenmayer System (L-system). The algorithm we use in \bricklayer\ is much simpler and second grade elementary school children\footnote{In the US, children in second grade are between the ages of 7 and 8.} have been able to follow this algorithm and build a Sierpinski gasket using actual LEGOs as shown in Figure \ref{fig-sierpinski-lego}. The algorithm we use is based on Pascal's triangle: odd numbers are mapped to one color of brick and even numbers are mapped to a different color of brick. The value at a particular position is computed from the two values above it. Our algorithm essentially tracks this information mod 2. First one places a 2x1 brick at the top of the triangle, then one successively processes rows lower in the triangle. The \bricklayer\ \emph{access} function is used to query the color of the two 1x1 bricks above the 2x1 brick currently being placed. If the colors of the above bricks agree, then the current brick to be placed, using the \bricklayer\ \emph{update} function, is assigned the even color; otherwise it is assigned the odd color. A \bricklayer\ implementation of this algorithm is shown in Appendix \ref{appendix-sierpinski}.

\section{Advanced Navigation}\label{section-advanced-navigation}

The philosophy behind the advanced navigation structure is to provide comprehensive access to the internals of \bricklayer. In addition to all the functionality described thus far, the AdvancedNavigation structure enables users to explicitly create and manipulate virtual space values directly. The rationale for this is still in a formative stage. For example, we are currently exploring how virtual spaces may be used to produce filter and lens-like effects. From a theoretical perspective, such functionality could be achieved at a lower level of abstraction, but we would argue that expression through lower-level functions constitutes an optimization of a concept that is more abstract.

Also provided are ``turtle graphics'' capabilities which enable fairly direct implementations of fractals which are specified in terms of Lindenmayer systems (L-systems). The Hilbert cube shown in Figure \ref{fourth} was created using the capabilities provided by the AdvancedNavigation structure. Depending on the target audience, the AdvancedNavigation structure may (probably should) be omitted from introductory course content.

\section{Conclusion}\label{section-conclusion}

The \bricklayer\ API connects SML programming with a rich domain of LEGO\textregistered\ artifacts. In \bricklayer, user developed predicates and brick functions characterize a class of fixed-width computations whose combination with \bricklayer\--provided generic traversals enable the creation of an incredibly broad range of artifacts. Within this realm of fixed-width computations, expressed as \bricklayer\ predicates and brick functions, problems whose solutions requiring sophisticated computational thinking abound.
Furthermore, a number of concepts have direct correspondences with topics taught in discrete math courses. For example, (1) the conjunction of predicates corresponds to set \emph{union}, (2) the outer conditions of nested conditionals can be used to restrict computational thinking to a particular domain of discourse, (3) computations involving division and modulus can be used to create repeating patterns as well as equivalence classes, and (4) triangles -- in their many shapes and orientations -- have strong conceptual analogies to nested loops. As a result, extensive meaningful study is possible within the realm of fixed-width computation.

In \bricklayer, recursive thinking is introduced gradually in the BasicNavigation structure. User-defined recursive functions whose behavior mirrors that of simple for-loops (e.g., a function that iterates over the z dimension of a virtual cube) can be composed with a variety of \bricklayer\ traversal functions (e.g., traverse the XY plane for a given value of z and apply a brick function to every point encountered) to create a broad range of artifacts. Simple recursion can then be composed to form nested loop functions gradually increasing the recursive complexity of an implementation. The end point is reached when \bricklayer\ traversal functions are no longer used to create structures. After such a point is reached, study can continue using the functionality provided by the AdvancedNavigation structure. Key features of the AdvancedNavigation structure include: (1) 3D turtle graphics, and (2) the treatment of the virtual space as a value. Using the functionality provided by the AdvancedNavigation structure it is possible to implement (1) L-systems, and (2) computational models such as cellular automata and even Turing machines.


\bibliographystyle{eptcs}
\bibliography{bricklayer}

\newpage
\appendix
\section{Signatures of the Predicate, and BasicNavigation Structures}\label{appendix-predicate}

\begin{lstlisting}
signature PRED =
sig
    type point         (* int * int * int *)
    type brick_type    (* an enumerated type whose values denote the bricks
                          supported in Bricklayer *)
    type cube_side     (* int *)
    type predicate     (* point -> bool *)

    val slideShow      : cube_side -> predicate -> brick_type -> unit
    val show           : cube_side * predicate * brick_type -> unit

    val xor            : cube_side * predicate * brick_type
                                   * predicate * brick_type -> unit

    val union          : cube_side * predicate * brick_type
                                   * predicate * brick_type * brick_type -> unit

    val intersection   : cube_side * predicate * predicate * brick_type -> unit
    val difference     : cube_side * predicate * predicate * brick_type -> unit
    val compliment     : cube_side * predicate * brick_type -> unit
end;
\end{lstlisting}

\begin{lstlisting}
signature BASIC_NAVIGATION =
sig
    type point          (* int * int * int *)
    type brick_type     (* an enumerated type whose values denote the bricks
                           supported in Bricklayer *)
    type brick_function (* point -> brick_type *)

    val build           : int * int * int -> unit
    val show            : string -> unit
    val slideShow       : unit -> unit

    val traverseWithin  : point -> point -> brick_function -> unit
    val traverseXYZ     : brick_function -> unit
    val traverseXY      : brick_function -> int -> unit
    val traverseXZ      : brick_function -> int -> unit
    val traverseYZ      : brick_function -> int -> unit

    val access          : point -> brick_type
    val update          : brick_type -> point -> unit

    val line            : point -> point -> brick_type -> unit
    val multiBrickLine  : point -> point -> brick_function -> unit

    val placeBrick      : int -> int -> int -> brick_type -> point -> unit
end;
\end{lstlisting}

\newpage
\section{Code for the Sierpinski Gasket}\label{appendix-sierpinski}

\begin{lstlisting}
fun sierpinski n = (* assumes n = 2^m + 1 *)
    let
        val max = n-1
        val mid = max div 2

        (* ------------------------------------------------------------------ *)
        fun aboveSame(x,y) =
            let
                val aboveLeft  = access(x,y+1,0)
                val aboveRight = access(x+1,y+1,0)
            in
                aboveLeft <> EMPTY andalso aboveLeft = aboveRight
            end

        (* ------------------------------------------------------------------ *)
        fun binaryUpdate brick (x,y)  =
            (
                update brick (x,y,0);
                update brick (x+1,y,0)
            )

        (* ------------------------------------------------------------------ *)
        fun drawRow (x,y) xHi =
            if x <= xHi then
                let
                    val brick = if aboveSame (x,y) then RED else BLUE
                in
                    binaryUpdate brick (x,y);
                    drawRow (x+2,y) xHi
                end
            else ()

        (* ------------------------------------------------------------------ *)
        fun processRows(lo, hi) =
            if hi < max then
                (
                    drawRow(lo,lo) hi;
                    processRows(lo-1,hi+1)
                )
            else ()

        (* ------------------------------------------------------------------ *)
    in
        build(n,n,1);

        processRows(mid, mid); 

        show "Sierpinski Gasket"
    end;
\end{lstlisting}

\end{document}